\documentclass[12pt]{article}
\usepackage{amsmath,amsfonts,feynmp,epsf}

\makeatletter\@addtoreset{equation}{section}\makeatother

\def\be{\begin{equation}}
\def\ee{\end{equation}}
\def\bea{\begin{eqnarray}}
\def\eea{\end{eqnarray}}

\newcommand{\m}{\mu}
\newcommand{\n}{\nu}

\makeatletter\@addtoreset{equation}{section}\makeatother

\def\ie{\begin{equation}\begin{aligned}}
\def\fe{\end{aligned}\end{equation}}

\renewcommand{\title}[1]{\vbox{\center\LARGE{#1}}\vspace{5mm}}
\renewcommand{\author}[1]{\vbox{\center#1}\vspace{5mm}}
\newcommand{\address}[1]{\vbox{\center\em#1}}
\newcommand{\email}[1]{\vbox{\center\tt#1}\vspace{5mm}}

\begin{document}

\unitlength = .8mm
\begin{titlepage}
\begin{center}
\hfill \\
\hfill \\
\vskip 1cm

\title{Families of Conformal Fixed Points of\\ ${\cal N}=2$ Chern-Simons-Matter Theories}

\author{Chi-Ming Chang$^{1,a}$ and
Xi Yin$^{1,2, b}$}

\address{
${}^1$Jefferson Physical Laboratory, Harvard University,\\
Cambridge, MA 02138 USA}
\address{
${}^2$School of Natural Sciences,
Institute for Advanced Study,\\
Princeton, NJ 08540 USA}

\email{$^a$cmchang@physics.harvard.edu,
$^b$xiyin@fas.harvard.edu}

\end{center}

\abstract{
We argue that a large class of ${\cal N}=2$ Chern-Simons-matter theories in three dimensions
have a continuous family of exact IR fixed points described by suitable quartic superpotentials, based on holomorphy. The entire family exists in the perturbative regime. A nontrivial check is performed by computing the 4-loop beta function of the quartic couplings, in the 't Hooft limit, with a large number of flavors. We find that the 4-loop beta function can only
deform the family of 2-loop fixed points, and does not change the dimension of this family.
We further present an explicit computation of a perturbative correction to the Zamolodchikov metric on this space of 
three-dimensional superconformal field theories.
}

\vfill

\end{titlepage}

\section{Introduction}

The Chern-Simons-matter (CSM) theories \cite{Schwarz:2004yj,Ivanov:1991fn,Gaiotto:2007qi,Avdeev:1991za,Avdeev:1992jt,Chen:1992ee,Kapustin:1994mt,Gaiotto:2008sd,Hosomichi:2008jd,Aharony:2008ug,Benna:2008zy} provide a large class of (super)conformal field theories in three-dimensions. It was pointed out in \cite{Gaiotto:2007qi} that even with given gauge group and matter content, the ${\cal N}=2$ CSM theory admits a large number of exact infrared fixed points, at least in the perturbative regime. In this paper, we make a very simple extension of the argument of \cite{Gaiotto:2007qi} to show that
in fact an entire continuous family of exact conformal fixed points exist; they are given by ${\cal N}=2$ CSM theory with appropriate quartic superpotentials. 

Our general argument will be based on holomorphy of the effective superpotential, and promoting superpotential coefficients to dynamical
chiral fields a la Seiberg \cite{Seiberg:1993vc}. An explicit 4-loop check will be performed. We find nontrivial cancelation of certain components of the 4-loop beta functions, consistent with the claim that the family of two-loop IR fixed points survive to all loop order (the precise RG fixed point locus may be deformed by higher loop effects).

While in a general ${\cal N}=2$ CSM theory, the $U(1)_R$ charge of the matter fields can be renormalized, there is no anomalous $U(1)_R$ charge along the continuous family of fixed points. With appropriate choices of matter content, one special point in this family is the ${\cal N}=3$ CSM theory. 
One moves along the family by turning on quartic chiral primary deformations of the superpotential.
On this space of superconformal CSM theories, there is a natural notion of metric -- the Zamolodchikov metric \cite{Zamolodchikov:1986gt}. We will consider the example of ${\cal N}=2$ $U(N)$ CSM theory with $M$ adjoint matter fields. At the leading nontrivial order, the family of fixed points modulo the $U(M)$ flavor symmetry (the quotient space denoted by ${\cal M}$) is given by a symplectic quotient of the linear complex vector space $V$ of all quartic superpotential coefficients, and the metric is the natural one associated with the symplectic form. We will compute the next-to-leading order perturbative correction to this metric in the 't Hooft limit. It will turn out that the moduli space ${\cal M}$ is a symplectic quotient defined by a deformed symplectic form on $V$.
However, the corrected Zamolodchikov metric, while still K\"ahler, is different from the one induced from the symplectic quotient.

In the next section, we will present the non-renormalization argument. Section 3 discusses the check via the 4-loop beta function, with details of the computation in Appendix A and B. While one may perform the computation using supergraph techniques, we found it more convenient to work with ordinary Feynman diagrams in component fields, utilizing the ``graphical rules" described in Appendix B. Section 4 studies the perturbative Zamolodchikov metric on the space of fixed points. The details of the computation of the metric are given in Appendix C and D. We summarize the results and conclude in section 5.

\section{A non-renormalization theorem}

Let us start by considering the example of ${\cal N}=2$ Chern-Simons-matter theory with $U(N)$ gauge group and $M$ {\sl adjoint} flavors. $k$ will denote the Chern-Simons level. For convenience, we will be mostly working in the 't Hooft limit, i.e. $N,k\to \infty$ with $\lambda=N/k$ fixed and treated perturbatively. This is a natural limit to consider, having in mind the holographic dual. Most of our arguments here can be straightforwardly generalized to finite $N$. The chiral matter superfields are denoted $\Phi_i$, with the flavor index $i=1,\cdots,M$. We will consider a general quartic (single trace) superpotential,
\ie\label{supw}
W = \frac{1}{4}\sum_{i,j,k,l} \alpha_{ijkl} {\rm Tr} (\Phi_i \Phi_j \Phi_k \Phi_l).
\fe
As argued in \cite{Gaiotto:2007qi}, the theory with $W=0$ is a superconformal field theory, in which the matter field $\Phi_i$ acquires a quantum corrected $U(1)_R$ charge, $J_\Phi={1\over 2}+{\cal O}({1\over k^2})$. Our normalization convention for the $U(1)_R$ charge $J$ is such that the unitarity bound on the scaling dimension of an operator of charge $J$ is $\Delta\geq J$. This bound is saturated by chiral primaries.
In the $W=0$ theory, every chiral operator is also a chiral primary. Nevertheless, the chiral primaries still acquire anomalous dimensions, which are equal to their anomalous $U(1)_R$ charges. Therefore, the operator ${\rm Tr} (\Phi_i \Phi_j \Phi_k \Phi_l)$ has dimension $4J_\Phi$ at the origin of the space of $\alpha$'s.
It was argued and also shown in explicit computation in \cite{Gaiotto:2007qi} that $J_\Phi<{1\over 2}$, i.e. $W=0$ is an unstable fixed point. Further, the beta function for $\alpha_{ijkl}$ (defined by normalizing the kinetic term for $\Phi_i$'s) up to two-loop order takes the form
\ie
\mu{d\alpha_{ijkl}\over d\mu} = (4J_\Phi-2) \alpha_{ijkl} + {1\over 4\pi^2}B_{(\underline{i}}{}^r \alpha_{r\underline{jkl})} + {\rm higher~loop}
\fe
where $(\underline{ijkl})$ stands for cyclic symmetrization. $B_i{}^j={1\over 2}N^2 \alpha_{iklm} \overline\alpha^{jklm}$ comes from the two-loop wave function renormalization in the corresponding Wess-Zumino model (obtained by decoupling the Chern-Simons gauge field).

Starting with the superpotential (\ref{supw}) in the UV, we can consider the Wilsonian effective action, of the form
\ie
S_{CS}^{{\cal N}=2}(V)+\int d^3x\int d^4\theta K(\Phi_i,\overline\Phi_i,V)
+ \int d^3x d^2\theta \sum  f_{ijkl}(k) \alpha_{ijkl} {\rm Tr} (\Phi_i \Phi_j \Phi_k \Phi_l) + c.c.
\fe
For now we are working in the 't Hooft limit, and hence the multi-trace operators are ignored in the effective action. Note however that our argument will go through even with multi-trace operators included. In particular, the effective superpotential only contains the quartic terms as in the classical superpotential. This follows from holomorphy of the effective superpotential in $\Phi_i$ and the $U(1)$ R-symmetry. Note that unlike in four-dimensional gauge theories \cite{Seiberg:1993vc, ArkaniHamed:1997mj}, here there is no anomaly in the global $U(1)$ symmetries, nor a dynamically generated scale, to allow for non-perturbatively generated superpotentials. Further, by promoting $\alpha_{ijkl}$ to dynamical chiral fields, one sees that the effective superpotential is also holomorphic in $\alpha_{ijkl}$.
%\footnote{It is important that we are dealing with Wilsonian rather than 1PI effective action here. In the %1PI effective action, there can be nonlocal terms in the K\"ahler potential so that after one substitutes %the expectation values of $\alpha_{ijkl}$ back in they look like superpotential terms.} 
By assigning an $R$-symmetry\footnote{This $R$-symmetry is not to be confused with the $U(1)_R$ of the superconformal algebra.} charge $2$ to $\alpha_{ijkl}$, $1$ to $\theta$ (and $-1$ to $\bar\theta$), and $0$ to $\Phi_i$'s,
we conclude that the effective superpotential must be linear in $\alpha_{ijkl}$, and that the superpotential coefficient can only be renormalized by the Chern-Simons coupling $1/k$.\footnote{One may worry about the linear mixing of $\alpha_{ijkl}$ with say $\alpha_{ilkj}$, which is consistent with the $U(M)$ flavor symmetry. However, this is not possible because ${\rm Tr}\Phi^4$ would be a chiral primary in {\sl the $W=0$ theory}, and therefore do not mix with one another at leading order in $\alpha$.}
As pointed out in \cite{Gaiotto:2007qi}, such corrections will occur in general, since one cannot promote the Chern-Simons level $k$ to a dynamical field without breaking gauge symmetry.

After normalizing the two-derivative kinetic term for $\Phi_i$ in the K\"ahler potential, we see that the quantum correction to $\alpha_{ijkl}$ amounts to an anomalous dimension for the operator $ {\rm Tr} (\Phi_i \Phi_j \Phi_k \Phi_l) $ together with a wave function renormalization. So, in fact, we expect
\ie
\mu{d\alpha_{ijkl}\over d\mu} = (4J_\Phi(k)-2) \alpha_{ijkl} + {1\over 4\pi^2}B_{(\underline{i}}{}^r \alpha_{r\underline{jkl})}
\fe
to hold exactly, for some $B_i{}^j(\alpha,\overline\alpha,k)={1\over 2}N^2 \alpha_{iklm} \overline\alpha^{jklm}+ ({\rm higher~ order~terms~in~}1/k,\alpha,\bar\alpha)$. Here $\alpha_{ijkl}$ are considered to be of the same order as $1/k$, as is the case along the $W\not=0$ two-loop fixed point loci. $J_\Phi(k)$ is the quantum corrected $U(1)_R$ charge of $\Phi_i$ in the $W=0$ theory, which is a function of $k$ only.\footnote{If we did not have the $U(M)$ flavor symmetry, of course, the $\Phi_i$'s may have different anomalous $U(1)_R$ charges depending on their representation content, in the $W=0$ theory.} We then conclude that the IR fixed points, up to the global flavor symmetry $U(M)$, is parameterized by the quotient space
\ie
{\cal M}=\left\{\alpha_{ijkl}:\,B_i{}^j = c\,\delta_i^j\right\}/U(M)
\fe
where $c=4\pi^2(2-4J_\Phi)>0$. Denote by $V$ the linear vector space of all $\alpha_{ijkl}$'s. ${\cal M}$ is generally a deformation of the standard symplectic quotient $V//U(M)$
by the 't Hooft coupling $\lambda=N/k$. We will revisit the geometry of ${\cal M}$ in section 4. When $M=2n$ is even, one point on ${\cal M}$ is given by the ${\cal N}=3$ CSM theory with $n$ adjoint hypermultiplets.

So in the perturbative regime, the IR fixed points of the ${\cal N}=2$ $U(N)$ CSM with $M$ adjoint matter fields are given by the $W=0$ fixed point together with the fixed point manifold ${\cal M}$ (up to the $U(M)$ flavor symmetry). The tangent directions of ${\cal M}$ are in 1-1 correspondence with quartic chiral primary operators.\footnote{More precisely, the tangent vectors along the fixed point loci in $V$ (before quotienting by $U(M)$) are a linear combination of the quartic chiral primaries and the scalar operators in the supermultiplet of the $U(M)$ flavor currents.} It then follows that the $U(1)_R$-charge of $\Phi_i$ along ${\cal M}$ is given exactly by $J={1\over 2}$, in contrast to $J_\Phi<{1\over 2}$ at $W=0$. When the number of flavors $M$ is even, the non-renormalization of $U(1)_R$ charge is well known at the ${\cal N}=3$ point on ${\cal M}$.  Now we conclude that this property continues to hold even when one deforms marginally away from the ${\cal N}=3$ point. At a given point on ${\cal M}$ with superpotential $W$, the chiral operators of the form ${\rm Tr}(\Phi_i \partial_j W)$ are descendants, and are transverse to the IR fixed point manifold in $V$. On the other hand, the explicit expressions of the quartic chiral primaries are dependent on $k$, and can be determined by looking at the tangent directions of ${\cal M}$. We will return to this in section 4.

Let us comment that the holomorphy argument above applies only to the Wilsonian effective action and not to the 1PI effective action \cite{ArkaniHamed:1997mj, Weinberg:1998uv}. This is because in the 1PI effective action, where massless modes are integrated out, nonlocal terms may be generated in the K\"ahler potential such that when one replaces the spurious chiral fields by their expectation values $\alpha_{ijkl}$, the term
looks like a superpotential term with non-holomorphic dependence on $\alpha_{ijkl}$ (see \cite{Poppitz:1996na}). In computing higher loop contributions to the beta function, the result from 1PI RG and Wilsonian RG may differ, depending on renormalization schemes. Nevertheless, the dimensionality of the loci of IR fixed points in the space of couplings $\alpha_{ijkl}$ clearly should not depend on the choice of renormalization group. The Chern-Simons gauge field may appear subtle from the perspective of Wilsonian RG. On one hand, the gauge fields are effectively infinitely massive and have no propagating degrees of freedom; on the other hand, they give rise to long range interactions, which may be thought of as a non-abelian generalization of the anyon statistics of the matter fields.
In principle, they must be treated carefully, using the regularization of \cite{Hayashi:1998ca} which cuts off the momenta in a way that
preserves supersymmetry and gauge symmetry manifestly. This is achieved, for instance, by replacing the Chern-Simons gauge super-propagator in ${\cal N}=2$ superspace by
\ie
\int_0^\infty d\tau f(\Lambda\tau) e^{-\tau D\bar D}\delta^7(z)
\fe
for some function $f(\tau)$ that vanishes (as well as its derivatives to all orders) at $\tau=0$, and approaches 1 at $\tau=\infty$. Here $\delta^7(z)\equiv \delta^3(x)\delta^4(\theta)$, and we did not take into account gauge fixing. If we are to integrate out a momenta shell from $\Lambda$ to $\Lambda+\delta\Lambda$, we may replace the regularized propagator by
\ie
\delta \Lambda \int_0^\infty d\tau \,\tau f'(\Lambda\tau) e^{-\tau D\bar D}\delta^7(z)
=-{\delta \Lambda\over\Lambda} \int_0^\infty d\tau f(\Lambda\tau) \partial_\tau\left[\tau e^{-\tau D\bar D}\right] \delta^7(z)
\fe
With the choice $f(\tau)=\theta(\tau-1)$, this is simply ${\delta\Lambda\over\Lambda^2} e^{-{D\bar D\over\Lambda}}\delta^7(z)$.

It is straightforward to generalize the non-renormalization argument to ${\cal N}=2$ CSM theories with any gauge groups and any matter representation content. The renormalization of the quartic superpotential coefficients, or Yukawa couplings, can be entirely absorbed into wave function renormalization. In particular, the a priori nontrivial $k$-dependent renormalization of the superpotential coefficients are entirely due to the anomalous dimensions of the chiral matter fields in the $W=0$ theory. The ``generic branch" of conformal fixed points are described by
\ie\label{gb}
B_i{}^j(\alpha,\overline\alpha,k) = 16\pi^2\left[ {1\over 2}-J_i(k) \right] \delta_i^j
\fe
where $B_i{}^j(\alpha,\bar\alpha,k)=(B_{WZ})_i{}^j(\alpha,\bar\alpha)+{\cal O}(1/k^2)$. Here $B_{WZ}$ represents the wave function renormalization in the corresponding Wess-Zumino model, and $J_i(k)$ is the quantum corrected $U(1)_R$ charge of the field $\Phi_i$ in the $W=0$ theory. For general matter content, there may also be non-generic branches of IR fixed points, where some of the $\alpha$'s set to zero, and (\ref{gb}) only needs to be satisfied for a subset of $\Phi_i$'s.

\section{A 4-loop check}

In the previous section we have given a holomorphy argument that the manifold of two-loop IR fixed points survives to all-loop order. That is, while the loci of the family of two-loop fixed points in $V$ may be deformed by higher loop effects, the dimension of the family remains unchanged. This is not at all obvious from the perspective of 1PI RG. In the 1PI effective action, a priori, there are terms that could potentially contribute to the beta function of $\alpha_{ijkl}$ in the form
\ie\label{poss}
\beta_{ijkl} = \beta_{ijkl}^{2-loop} + {C\over k^2} \alpha_{(\underline{ij}mn}
\bar\alpha^{mnpq}\alpha_{pq\underline{kl})} + \cdots
\fe
where we exhibited one possible 4-loop contribution. $C$ is a constant coefficient that generally depends on $M$ and $N$. Such a 4-loop contribution cannot be absorbed into the wave function renormalization of the matter fields. If $C$ is nonzero, the family of two-loop fixed points will further flow to a submanifold of lower dimension (possibly discrete points).
While the higher loop beta function in 1PI RG may not agree with that of the Wilsonian RG in general, the dimensionality of the loci of IR fixed points should not depend on which RG we use. Therefore we expect the higher-than-two-loop contributions such as the second term on the RHS of (\ref{poss}) to vanish. We will now check this explicitly
at 4-loop order.

We will make a few simplifying assumptions. Firstly, we shall work at the planar level. We expect the same conclusion to hold with non-planar diagrams included as well, but the computation would be more complicated. (Having in mind the holographic dual, the planar limit is interesting on its own.) Secondly, we will take the number of flavors, $M$, to be parametrically large, and start by looking at the leading nontrivial contribution in the large $M$ limit. This reduces the number of diagrams drastically. The subleading $1/M$ contributions involve many more diagrams, whose explicit computations are not consider in the current paper.

At the planar level, the potential contributions to the effective superpotential that cannot be absorbed into wave function renormalization take the general form
\begin{equation}
(c_1 M + c_2) {4\pi^2\over k^2}N^4\alpha_{ijmn}\bar\alpha^{mnpq}\alpha_{pqkl}
{\rm Tr}\left(\Phi^i\Phi^j\Phi^k\Phi^l\right)
\label{effective_superpotential}
\end{equation}
where $c_1$ and $c_2$ are constants. To see this, let us examine the diagrams. While we will perform the computation using ordinary Feynman diagrams in component fields, it is convenient to organize them using ${\cal N}=2$ supergraphs. Our notation is explained in Appendix A. In a supergraph that contributes to 
(\ref{effective_superpotential}), the F-term vertices are contracted according to the following structure:

\bigskip

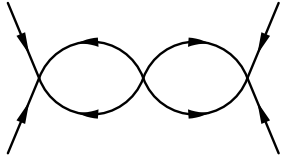
\centerline{\begin{fmffile}{basic}
        \begin{tabular}{c}
            \begin{fmfgraph*}(45,25)
                \fmfstraight
                \fmfset{arrow_len}{.3cm}\fmfset{arrow_ang}{12}
                \fmfleft{i1,i2}
                \fmfright{o1,o2}
                \fmf{fermion}{i2,v1}
                \fmf{fermion}{i1,v1}
                \fmf{fermion}{o1,v2}
                \fmf{fermion}{o2,v2}
                \fmf{fermion,right=.7,tension=.3}{v3,v1}
                \fmf{fermion,right=.7,tension=.3}{v3,v2}
                \fmf{fermion,left=.7,tension=.3}{v3,v1}
                \fmf{fermion,left=.7,tension=.3}{v3,v2}
            \end{fmfgraph*}
        \end{tabular}
\end{fmffile}}

%\bigskip
\noindent The non-abelian ${\cal N}=2$ CS action minimally coupled to chiral fields can be written in
${\cal N}=2$ superspace as
\ie
S_D = \int d^3 x \int d^4\theta \left\{ {k\over 2\pi}\int_0^1 dt {\rm Tr} \left[
V \bar D^\alpha \left( e^{-tV}D_\alpha e^{tV} \right) \right] + \sum_i \bar\Phi_i e^V \Phi_i \right\}
\fe
In Wess-Zumino gauge, the D-term supervertices involve the cubic interactions of the super gauge fields and the standard minimal coupling to matter fields, as described in Appendix A. These D-supervertices can be attached to the above graph to form a 4-loop diagram that contribute to the beta function. Some examples are
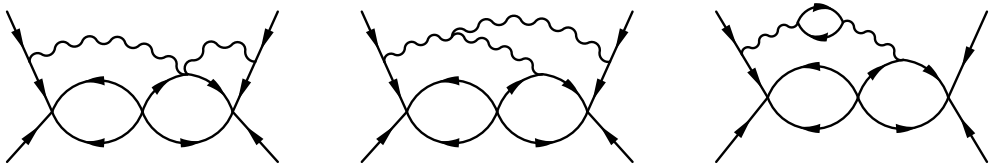
\begin{equation}\nonumber
\begin{array}{ccc}
\begin{fmffile}{ex1}
        \begin{tabular}{c}
            \begin{fmfgraph*}(45,25)
                \fmfstraight
                \fmfset{arrow_len}{.3cm}\fmfset{arrow_ang}{12}
                \fmfleft{i1,i2}
                \fmfright{o1,o2}
                \fmf{fermion}{i2,v4}
                \fmf{fermion}{v4,v1}
                \fmf{fermion}{i1,v1}
                \fmf{fermion}{o1,v2}
                \fmf{fermion}{o2,v6}
                \fmf{fermion}{v6,v2}
                \fmffixed{(.6h,0)}{v1,v3}
                \fmffixed{(.6h,0)}{v3,v2}
               %\fmffixed{(-.3h,.25h)}{v3,v4}
                \fmffixed{(.3h,.25h)}{v3,v5}
                \fmf{fermion,right=.7,tension=.3}{v3,v1}
                \fmf{fermion,right=.7,tension=.3}{v3,v2}
                \fmf{fermion,left=.7,tension=.3}{v3,v1}
                \fmf{fermion,left=.3,tension=.3}{v3,v5,v2}
                \fmf{photon,left=.35,tension=0}{v4,v5}
                \fmf{photon,left=.7,tension=0}{v5,v6}
             \end{fmfgraph*}
        \end{tabular}
        \end{fmffile}
&~~~~
\begin{fmffile}{ex2}
        \begin{tabular}{c}
            \begin{fmfgraph*}(45,25)
                \fmfstraight
                \fmfset{arrow_len}{.3cm}\fmfset{arrow_ang}{12}
                \fmfleft{i1,i2}
                \fmfright{o1,o2}
                \fmf{fermion}{i2,v4}
                \fmf{fermion}{v4,v1}
                \fmf{fermion}{i1,v1}
                \fmf{fermion}{o1,v2}
                \fmf{fermion}{o2,v6}
                \fmf{fermion}{v6,v2}
                \fmffixed{(.6h,0)}{v1,v3}
                \fmffixed{(.6h,0)}{v3,v2}
               %\fmffixed{(-.3h,.25h)}{v3,v4}
                \fmffixed{(.3h,.25h)}{v3,v5}
                \fmffixed{(-.3h,.5h)}{v3,v7}
                \fmf{fermion,right=.7,tension=.3}{v3,v1}
                \fmf{fermion,right=.7,tension=.3}{v3,v2}
                \fmf{fermion,left=.7,tension=.3}{v3,v1}
                \fmf{fermion,left=.3,tension=.3}{v3,v5,v2}
                \fmf{photon,left=.17,tension=0}{v4,v7,v5}
                \fmf{photon,left=.35,tension=0}{v7,v6}
             \end{fmfgraph*}
        \end{tabular}
        \end{fmffile}
&~~~~
\begin{fmffile}{ex3}
        \begin{tabular}{c}
            \begin{fmfgraph*}(45,25)
                \fmfstraight
                \fmfset{arrow_len}{.3cm}\fmfset{arrow_ang}{12}
                \fmfleft{i1,i2}
                \fmfright{o1,o2}
                \fmf{fermion}{i2,v4}
                \fmf{fermion}{v4,v1}
                \fmf{fermion}{i1,v1}
                \fmf{fermion}{o1,v2}
                \fmf{fermion}{o2,v2}
                \fmffixed{(.6h,0)}{v1,v3}
                \fmffixed{(.6h,0)}{v3,v2}
               %\fmffixed{(-.3h,.25h)}{v3,v4}
                \fmffixed{(.3h,.25h)}{v3,v7}
                \fmffixed{(-.4h,.5h)}{v3,v5}
                \fmffixed{(-.1h,.5h)}{v3,v6}
                \fmf{fermion,right=.7,tension=.3}{v3,v1}
                \fmf{fermion,right=.7,tension=.3}{v3,v2}
                \fmf{fermion,left=.7,tension=.3}{v3,v1}
                \fmf{fermion,left=.3,tension=.3}{v3,v7,v2}
                \fmf{photon,left=.1,tension=0}{v4,v5}
                \fmf{fermion,left=.7,tension=0}{v5,v6,v5}
                \fmf{photon,left=.1,tension=0}{v6,v7}
             \end{fmfgraph*}
        \end{tabular}
        \end{fmffile}
\end{array}
\end{equation}
In the limit of large $M$, the third diagram dominates the first two, due to the factor $M$ coming from the matter-loop-corrected vector superfield propagator. For now, we will consider this limit and ignore diagrams such as the first two above. In other words, we will be computing $c_1$ but not $c_2$ in (\ref{effective_superpotential}).
There are only three types of planar 4-loop supergraphs that contribute to $c_1$, 
given by
\begin{equation}
\begin{array}{ccc}
\begin{fmffile}{A}
        \begin{tabular}{c}
            \begin{fmfgraph*}(45,25)
                \fmfstraight
                \fmfset{arrow_len}{.3cm}\fmfset{arrow_ang}{12}
                \fmfleft{i1,i2}
                \fmfright{o1,o2}
                \fmf{fermion}{i2,v1}
                \fmf{fermion}{i1,v1}
                \fmf{fermion}{o2,v2}
                \fmf{fermion}{o1,v2}
                \fmffixed{(.6h,0)}{v1,v3}
                \fmffixed{(.6h,0)}{v3,v2}
               \fmffixed{(-.3h,.25h)}{v3,v4}
                \fmffixed{(.3h,.25h)}{v3,v7}
                \fmffixed{(-.15h,.5h)}{v3,v5}
                \fmffixed{(.15h,.5h)}{v3,v6}
                \fmf{fermion,right=.3,tension=.3}{v3,v4,v1}
                \fmf{fermion,right=.7,tension=.3}{v3,v2}
                \fmf{fermion,left=.7,tension=.3}{v3,v1}
                \fmf{fermion,left=.3,tension=.3}{v3,v7,v2}
                \fmf{photon,left=.3,tension=0}{v4,v5}
                \fmf{fermion,left=.7,tension=0}{v5,v6,v5}
                \fmf{photon,left=.3,tension=0}{v6,v7}
             \end{fmfgraph*}
        \end{tabular}
        \end{fmffile}
&~~~~
\begin{fmffile}{B}
        \begin{tabular}{c}
            \begin{fmfgraph*}(45,25)
                \fmfstraight
                \fmfset{arrow_len}{.3cm}\fmfset{arrow_ang}{12}
                \fmfleft{i1,i2}
                \fmfright{o1,o2}
                \fmf{fermion}{i2,v4}
                \fmf{fermion}{v4,v1}
                \fmf{fermion}{i1,v1}
                \fmf{fermion}{o1,v2}
                \fmf{fermion}{o2,v2}
                \fmffixed{(.6h,0)}{v1,v3}
                \fmffixed{(.6h,0)}{v3,v2}
               %\fmffixed{(-.3h,.25h)}{v3,v4}
                \fmffixed{(.3h,.25h)}{v3,v7}
                \fmffixed{(-.1h,.5h)}{v3,v6}
                \fmffixed{(-.4h,.5h)}{v3,v5}
                \fmf{fermion,right=.7,tension=.3}{v3,v1}
                \fmf{fermion,right=.7,tension=.3}{v3,v2}
                \fmf{fermion,left=.7,tension=.3}{v3,v1}
                \fmf{fermion,left=.3,tension=.3}{v3,v7,v2}
                \fmf{photon,left=.1,tension=0}{v4,v5}
                \fmf{fermion,left=.7,tension=0}{v5,v6,v5}
                \fmf{photon,left=.1,tension=0}{v6,v7}
             \end{fmfgraph*}
        \end{tabular}
        \end{fmffile}
&~~~~
\begin{fmffile}{C}
        \begin{tabular}{c}
            \begin{fmfgraph*}(45,25)
                \fmfstraight
                \fmfset{arrow_len}{.3cm}\fmfset{arrow_ang}{12}
                \fmfleft{i1,i2}
                \fmfright{o1,o2}
                \fmf{fermion}{i2,v4}
                \fmf{fermion}{v4,v1}
                \fmf{fermion}{i1,v1}
                \fmf{fermion}{v7,v2}
                \fmf{fermion}{o2,v7}
                \fmf{fermion}{o1,v2}
                \fmffixed{(.6h,0)}{v1,v3}
                \fmffixed{(.6h,0)}{v3,v2}
                \fmffixed{(-.15h,.5h)}{v3,v5}
                \fmffixed{(.15h,.5h)}{v3,v6}
                \fmf{fermion,right=.7,tension=.3}{v3,v1}
                \fmf{fermion,right=.7,tension=.3}{v3,v2}
                \fmf{fermion,left=.7,tension=.3}{v3,v1}
                \fmf{fermion,left=.7,tension=.3}{v3,v2}
                \fmf{photon,left=.1,tension=0}{v4,v5}
                \fmf{photon,left=.1,tension=0}{v6,v7}
                \fmf{fermion,left=.7,tension=0}{v5,v6,v5}
             \end{fmfgraph*}
        \end{tabular}
        \end{fmffile}
\label{ABC}
\\
(a) & (b) & (c)
\end{array}
\end{equation}
Let us note that planarity forbids contributions to terms proportional to
$\alpha_{ijmn}\bar\alpha^{mpnq}\alpha_{pqkl}$ or $\alpha_{ijmn}\bar\alpha^{mqnp}\alpha_{pqkl}$ in the beta function. Similarly, in the beta function for $\alpha_{ijkl}$, planarity only allows 
the term $\alpha_{ijmn}\bar\alpha^{mnpq}\alpha_{pqkl}$ with the indices $\{i,j,k,l\}$ appearing in cyclic order.

In component fields, the super gauge field propagator involves the vector gauge field as well as the auxiliary fields $D,\sigma,\chi$. The latter can be integrated out to give quartic scalar-fermion vertices and sextic scalar vertices (not used here). These diagrams are computed explicitly in Appendix A. The coefficient $c_1$ is
\ie
c_1 = {1\over 2}(a+b+c)
\fe
where $a$, $b$, $c$ are constants computed from the three supergraphs $(a), (b), (c)$ listed above. We find
\ie
a=c={1\over 256\pi^2},~~~b=-{1\over 128\pi^2},
\fe
and they indeed sum up to zero. We note however that the individual supergraph contribution does not vanish, and so $c_1=0$ here is a consequence of cancellation among different supergraphs in the 1PI RG.

\section{The metric on ${\cal M}$}

The manifold ${\cal M}$, defined as the $W\not=0$ IR fixed points in $V$ modulo $U(M)$ flavor symmetry, is naturally equipped with a Zamolodchikov metric. The metric is defined by the coefficient of the two-point function of quartic chiral primaries and their conjugates that parameterize the tangent directions of ${\cal M}$. In particular, the geometry of ${\cal M}$ will generally depend on the 't Hooft coupling $\lambda$. We expect the generic CSM theory to have a holographic dual, which may or may not have a gravity limit at strong 't Hooft coupling.\footnote{The theories with a large number of adjoint flavors, in particular, are expected to only have a stringy holographic dual. This is because the number of chiral primaries grow exponentially with the dimension, which cannot happen in a supergravity theory compactified to $AdS_4$.} Heuristically, had there been a gravity dual say of the form M-theory on $AdS_4\times M_7$, $M_7$ being the base of a Calabi-Yau 4-fold cone, then the analog of the manifold ${\cal M}$ at
strong 't Hooft coupling would be the moduli space of this CY 4-fold cone.\footnote{See \cite{Martelli:2008a, Martelli:2008b, Martelli:2009, Hanany:2008, Franco:2008} for recent work on such theories.} The geometry of ${\cal M}$ at
strong 't Hooft coupling is difficult to understand from the field theory perspective. In this section, we will investigate the perturbative corrections to the Zamolodchikov metric on ${\cal M}$.

We may write the $W\not=0$ IR fixed point locus as
\ie\label{mpc}
\mu_i{}^j(\alpha,\overline\alpha,k) = r(k)\delta_i^j
\fe
Here $\mu_i{}^j$ is proportional to $B_i{}^j$, with a possibly $k$-dependent normalization factor for later convenience. So, $r(k)$ is not necessarily the same as ${1\over 2}-J(k)$. Up to two-loop contribution, and in the limit of large $M$, we have
$r(k)=M({4\pi N\over k})^2 +\cdots$. Its precise $k$ dependence is not important for our purpose.
The tangent directions $\delta\alpha_{ijkl}=c_{ijkl}$ are determined by
\ie
c_{mnpq} {\partial\over\partial\alpha_{mnpq}} \mu_i{}^j(\alpha,\overline\alpha,k)=0,~~~\forall i,j.
\fe
The quartic chiral primaries are then given by ${\cal O}_c=\sum c_{mnpq} {\rm Tr}(\phi_m \phi_n \phi_p \phi_q)$ for such $c$. The two-point functions of a quartic chiral primary and an anti-chiral primary, in the SCFT corresponding to a point on ${\cal M}$, take the form
\ie
\langle {\cal O}_c(x) \overline{\cal O}_{c'}(0)\rangle = {g(c,\bar c')\over |x|^4}.
\fe
where the coefficient $g(c,\bar c')$ is the Zamolodchikov metric.

To begin, let us examine the IR manifold ${\cal M}$ at two-loop order. The leading contribution to the Zamolodchikov metric is simply given by the free correlator,
\ie
g^{(0)}(c,\bar c') = \sum c_{mnpq}\bar c'^{mnpq}.
\fe
It is the standard Euclidean metric on the space of $\alpha_{ijkl}$'s, corresponding to the symplectic form
\ie
\omega^{(0)} = d\alpha_{mnpq}\wedge d\bar\alpha^{mnpq}.
\fe
The metric on ${\cal M}$ at the leading order is the one induced from the symplectic quotient by
the flavor symmetry $U(M)$.

At the next-to-leading order, we must consider the 4-loop-corrected IR manifold ${\cal M}$, as well as the two-loop contributions to the two-point functions of the quartic chiral primaries. With the 4-loop contributions taken into account, $\mu_i{}^j$ takes the form
\ie\label{mucorr}
\mu_i{}^j = N^2 \alpha_{imnp}\bar\alpha^{jmnp} +  a_1 N^4 \left(\alpha_{imnk}\bar\alpha^{mnpq}\alpha_{pqrs}\bar\alpha^{rskj}+\alpha_{kmni}\bar\alpha^{mnpq}\alpha_{pqrs}\bar\alpha^{rsjk}\right) + {\rm higher~ order}.
\fe
Here the constant $a_1$ is simply given by the 4-loop wave function renormalization in the Wess-Zumino model. All other 4-loop corrections to $B_i{}^j$ will be proportional to the first term in (\ref{mucorr}) and are subleading in $1/k^2$; they can be absorbed by a rescaling of $\mu_i{}^j$ and $r(k)$ in (\ref{mpc}).
$\mu_i{}^j$ is in fact a moment map associated with the symplectic form
\ie
\omega = N^2 d\alpha_{mnpq}\wedge d\bar\alpha^{mnpq}
+4a_1 N^4 \alpha_{rsij}\bar\alpha^{pqij} d\alpha_{mnpq}\wedge d\bar\alpha^{mnrs} + {\rm higher~ order~ terms}.
\label{a1}
\fe
By definition, given the $U(M)$ action as a vector field on $V$
\ie
v_i{}^j = \alpha_{imnp} {\partial\over\partial\alpha_{jmnp}} 
- \bar\alpha^{jmnp} {\partial\over\partial \bar\alpha^{imnp}},
\fe
we have $d\mu_i{}^j = \iota_{v_i{}^j}\omega$, where $\iota$ stands for the contraction with a vector field.

Generally, the two-loop correction to the two-point function of the quartic chiral primaries takes the form
\ie
g(c,\bar c') = f(k) c_{mnpq}\bar c'^{mnpq} + a_2 N^2 c_{ijmn} \bar\alpha^{mnpq}\alpha_{pqrs}\bar c'^{rsij}
+ a_3 N^2 c_{imnp} \bar\alpha^{mnpq}\alpha_{qrst}\bar c'^{rsti},\label{a2}
\fe
where $f(k) = 1+ {\cal O}(1/k^2)$, $a_2,a_3$ are constants. $c_{ijkl}$ is constrained to be tangent to the IR manifold, in particular,
$c_{imnp}\bar\alpha^{mnpq} = {\cal O}(c/k^3)$ from the two-loop constraints. So we can ignore $a_3$ in the expression for $g(c,\bar c')$. The coefficient $a_1$ is computed in Appendix C, and $a_2$ is computed in Appendix D. We find that
\ie
a_1 = -{1\over 128},~~~~a_2=-{1\over 16}.
\fe
Therefore, the Zamolodchikov metric with the next-to-leading correction included takes the form
\ie\label{gcck}
g(c,\bar c') = f(k) c_{mnpq}\bar c'^{mnpq} -{N^2\over 16} c_{ijmn} \bar\alpha^{mnpq}\alpha_{pqrs}\bar c'^{rsij} + \cdots
\fe
whereas $\alpha_{ijkl}$'s are constrained by
\ie\label{momt}
\mu_i{}^j &= N^2 \alpha_{imnp}\bar\alpha^{jmnp} -{N^4\over 128} \left(\alpha_{imnk}\bar\alpha^{mnpq}\alpha_{pqrs}\bar\alpha^{rskj}+\alpha_{kmni}\bar\alpha^{mnpq}\alpha_{pqrs}\bar\alpha^{rsjk}\right) +
\cdots
\\
&= r(k) \delta_i{}^j
\fe
We note that due to a factor of 2 difference in the second term, $g(c,\bar c')$ is not the same as the natural symplectic metric on the quotient space ${\cal M}=V//U(M)$ defined using the symplectic form $\omega$. Nevertheless, $g(c,\bar c')$ is the restriction of a K\"ahler metric on the ambient space $V$ to the level set $\mu^{-1}(r(k)\delta_i^j)$; it follows easily that the Zamolodchikov metric on ${\cal M}$ is also K\"ahler, at least to the order we have computed, even though it is not the same as the K\"ahler metric induced from the symplectic quotient. We will now sketch an argument that the metric on ${\cal M}$ is K\"ahler to all order. 

The variation of the metric along a tangent direction corresponding to a chiral primary ${\cal O}_{c''}$ is
\ie\label{delg}
\delta_{c''} g(c,\bar c') = |x|^4 \left\langle {\cal O}_{c}(x) \left[ \int d^3y Q^2\cdot {\cal O}_{c''}(y)\right]\overline {\cal O}_{\bar c'}(0) \right\rangle
\fe 
The statement that $g$ is K\"ahler amounts to $\delta_{c_1} g(c_2,\bar c_3)=\delta_{c_2} g(c_1,\bar c_3)$, for chiral primaries $c_1,c_2,c_3$. Let us study the correlation function
\ie
F(c_1,c_2,\bar c_3;x,y,z) = \langle {\cal O}_{c_1}(x) \left[ Q^2\cdot {\cal O}_{c_2}(y) \right] \overline {\cal O}_{c_3}(z) \rangle
\fe
$Q^2\cdot {\cal O}_{c_2}$ is a primary with respect to the bosonic conformal algebra. Apart from potential contact terms, the spatial dependence of the three-point function of the primaries as above are fixed by conformal symmetry, to be $|x-y|^{-3}|y-z|^{-3} |x-z|^{-1}$. On the other hand, by Ward identity, we can move $Q^2$ from acting on ${\cal O}_{c_2}(y)$ to acting on ${\cal O}_{c_1}(x)$, and conclude that $F(c_1,c_2,\bar c_3;x,y,z) = F(c_2,c_1,\bar c_3;y,x,z)$. This is inconsistent with the naive spatial dependence determined by conformal symmetry, which implies that $F$ must vanish up to contact terms. 

Indeed, there are such contact terms. We can explicitly compute $F(c_1,c_2,\bar c_3;x,y,z)$ in perturbation theory. To leading nontrivial order, this is computed by the same diagrams as in Appendix D, interpreted as a three point function rather than a two-point function. It is given by
\ie
&F(c_1,c_2,\bar c_3;x,y,z) 
\\&\sim (c_1)_{ijkl}(c_2)_{mnpq}\bar c_3^{ijmn}\bar\alpha^{klpq}
{1\over |x-z|^2|y-z|^2}\int d^3w {1\over |x-w|^2 |y-w|^4}
+ {\rm higher~order}.
\fe
The integration over $w$ naively gives zero by analytic continuation in the exponents of the propagators. If we first integrate the integrand multiplied by a generic function of $x$ over $x$, and then integrate over $w$, we see that
\ie
\int d^3w {1\over |x-w|^2 |y-w|^4} = -{\pi\over 4}\delta^3(x-y).
\fe
This gives the contact term in $F(c_1,c_2,\bar c_3;x,y,z)$. With higher order contributions included, on dimensional grounds we expect $F$ to take the form
\ie
F(c_1,c_2,\bar c_3;x,y,z) = {f(c_1,c_2,\bar c_3)\over |x-z|^4}\delta^3(x-y).
\fe
where $f(c_1,c_2,\bar c_3)$ is symmetric in $c_1$ and $c_2$ by the Ward identity argument above. The closure of the K\"ahler form associated with $g(c,\bar c')$ then follows.

Coming back to (\ref{momt}), we may also perform a nonlinear redefinition of the coupling $\alpha_{ijkl}$,
\ie
\tilde\alpha_{ijkl} &= \alpha_{ijkl} - {N^2\over 256}\left(\alpha_{limn}\bar\alpha^{mnpq}\alpha_{pqjk}
+ \alpha_{ijmn}\bar\alpha^{mnpq}\alpha_{pqkl} \right)
\\
&= \alpha_{ijkl} - {N^2\over 128}\alpha_{(\underline{ij}mn}\bar\alpha^{mnpq}\alpha_{pq\underline{kl})}
\fe
where $(\underline{ij}\cdots\underline{kl})$ stands for cyclic symmetrization on the indices $\{i,j,k,l\}$, such that the moment map reduces to the standard one
\ie
\mu_i{}^j(\tilde\alpha,\bar{\tilde\alpha}) = N^2\tilde\alpha_{imnp}\bar{\tilde \alpha}^{jmnp} + (\rm{6-loop ~and~higher~order})
\fe
The tangent space basis vectors are now modified to
\ie
\tilde c_{ijkl} = c_{ijkl} - {N^2\over 128}\left[2\alpha_{(\underline{ij}mn}\bar\alpha^{mnpq}c_{pq\underline{kl})} + \alpha_{(\underline{ij}mn}\bar c^{mnpq}\alpha_{pq\underline{kl})} \right]
\fe
They satisfy $\tilde c_{imnp} \bar{\tilde \alpha}^{jmnp} = 0$ up to 6-loop contributions. The metric in the new coordinate system is written
\ie
&g(\tilde c,\bar{\tilde c};\tilde c',\bar{\tilde c}')= f(k) \tilde c_{ijkl}\bar{\tilde c}'^{ijkl}\\
&~-{N^2\over 32}\left( \tilde c_{ijmn} \bar\alpha^{mnpq}\alpha_{pqrs}{\bar{\tilde c}}'^{rsij}
-{1\over 4} \alpha_{ijmn}\alpha_{pqkl} \bar{\tilde c}^{ijkl} \bar{\tilde c}'^{mnpq}
-{1\over 4} \bar\alpha^{ijmn}\bar\alpha^{pqkl} {\tilde c}_{ijkl} {\tilde c'}_{mnpq} \right)+\cdots.
\fe
Due to the non-holomorphic change of coordinates, the metric is not Hermitian in this new coordinate system on ${\cal M}$.

So far we have focused on the chiral primaries ${\cal O}_c = c_{ijkl}{\rm Tr} \phi_i \phi_j \phi_k \phi_l$
which give rise to deformations along the IR fixed point loci $\mu^{-1}(r(k) \delta_i^j)\subset V$. In addition, there are the $U(M)$ rotation on the $\alpha_{ijkl}$'s, corresponding to the tangent vectors $v_i{}^j$. As exactly marginal deformations of the action, they can be written in terms of the operators
\ie
{\cal V}_i{}^j &= i \int d^2\theta {\rm Tr}\left(\Phi^j {\partial W\over\partial\Phi^i}\right) + c.c. \\
&= i \int d^2\theta \sum_{m,n,p}\alpha_{imnp}{\rm Tr} (\Phi^j \Phi^m \Phi^n \Phi^p) + c.c.
\fe
The ${\cal V}_i{}^j$'s lie in the same supermultiplet as the $U(M)$ flavor current.

Let us examine the quartic non-primary chiral operators ${\cal O}_i{}^j ={\rm Tr}\left[\phi^j {\partial \over\partial\phi^i}W(\phi)\right]$ more closely. While classically it is a descendant, the precise form of the descendant operator receives quantum corrections. This can be seen as follows. In general, ${\cal O}_i{}^j$ may be the linear combination of a purely descendant operator $\tilde{\cal O}_i{}^j$ with a chiral primary. $\tilde{\cal O}_i{}^j$ is the one orthogonal to all quartic chiral primaries. Namely, the corresponding tangent vector of the IR manifold $\tilde v_i^j$ satisfies
\ie\label{gvc}
g(\tilde v_i{}^j, \bar c) = 0
\fe
for all chiral primaries ${\cal O}_c$. On the other hand, since ${\cal O}_c$ is tangent to the level set of $\mu_i{}^j$, we have
\ie
\omega(v_i{}^j,\bar c) = 0.
\fe
We have seen that the next-to-leading order perturbative correction to $g$ does not agree with that of $\omega$, which implies that $\tilde v_i{}^j$ is different from $v_i{}^j$. In fact, demanding (\ref{gvc}) gives
\ie\label{otild}
\tilde {\cal O}_i{}^j &= {\cal O}_i{}^j +{N^2\over 32} \delta_{(\underline{k}}{}^j \alpha_{i\underline{lmn})} \bar \alpha^{mnpq}\alpha_{pqrs} {\rm Tr} \left(\phi^k\phi^l\phi^r\phi^s \right) + {\rm higher~order}
\\
&= {\cal O}_i{}^j +{N^2\over 128}  \left[\alpha_{i kmn} \bar \alpha^{mnpq}\alpha_{pqrs} {\rm Tr} \left(\phi^j\phi^k\phi^r\phi^s \right) + \alpha_{kimn} \bar \alpha^{mnpq}\alpha_{pqrs} {\rm Tr} \left(\phi^k\phi^j\phi^r\phi^s \right) \right.\\
&~~~\left.+\alpha_{i mkl} \bar \alpha^{jmpq}\alpha_{pqrs} {\rm Tr} \left(\phi^k\phi^l\phi^r\phi^s \right) + \alpha_{mikl} \bar \alpha^{mjpq}\alpha_{pqrs} {\rm Tr} \left(\phi^k\phi^l\phi^r\phi^s \right)\right] + {\rm higher~order}.
\fe
On the other hand, ${\cal V}_i{}^j = i Q^2 {\cal O}_i{}^j - i\overline Q^2 \overline{\cal O}_i{}^j$ is a descendant of the flavor current and is therefore orthogonal to all quartic chiral primaries. This is not in conflict with (\ref{otild}), since the descendant of the quartic chiral primary in $iQ^2{\cal O}_i{}^j$ may be canceled by a descendant of $\overline{\cal O}_i{}^j$. A more detailed investigation of the operator spectrum of this family of SCFTs is left to future work.

\section{Discussion}

We have argued that there are large classes of continuous families of three-dimensional ${\cal N}=2$ superconformal field theories, described by Chern-Simons-matter theories with appropriate quartic superpotentials. Such SCFTs in the large $N$ limit are expected to have holographic duals
as string theories in $AdS_4$. The best understood example is the theory of ABJM \cite{Aharony:2008ug}, which is dual to type IIA string theory on $AdS_4\times \mathbb {CP}^3$. However, all current examples of CSM theories with known or conjectured gravity duals have a parametrically small number of matter flavors (given the gauge group). As was pointed out in \cite{Gaiotto:2007qi}, this is because for a large number of flavors $M$, the number of chiral primaries will grow exponentially in their dimensions, which is faster than the growth of KK modes and is characteristic of string modes. For instance, suppose the superpotential $W$ is a homogeneous polynomial of degree $d$ in $M$ adjoint flavors $\Phi_i$, $i=1,\cdots, M$. The chiral operators that are descendants of primaries have the form ${\rm Tr} (f_i(\Phi) \partial_i W )$, where $\partial_i$ stands for the derivatives with respect to $\Phi_i$. For single trace operators of length $L$, the number of such descendants grow with $M$ like $M^{L-d+2}$, whereas the total number of chiral operators of length $L$ grow like 
$M^L/L$ at large $M$. From the perspective of the $AdS_4$ dual, this is a Hagedorn-like growth of states with Hagedorn temperature $T\sim (R \log M)^{-1}$. 
%It would be interesting to identify these dual string theories.

A nice feature of our family of CSM superconformal field theories is that, various features of SCFTs can be studied in perturbation theory.
In particular, we considered the perturbative corrections to the geometry of the moduli space\footnote{Here the moduli space refers to that of CFTs, not to be confused with the moduli space of vacua in a given CFT.} ${\cal M}$ of ${\cal N}=2$ CSM SCFTs, given by $U(N)$ Chern-Simons theory coupled to $M$ adjoint chiral matter fields and quartic superpotentials. At the leading nontrivial (two-loop) order, ${\cal M}$ is the symplectic quotient $V//U(M)$ defined by the standard symplectic form on $V$. This symplectic form, and hence the moduli space ${\cal M}=V//U(M)$, is deformed by 4-loop corrections. We also found a nontrivial 4-loop correction to the Zamolodchikov metric on ${\cal M}$, which is K\"ahler but is {\sl not} the same as the induced metric from the symplectic quotient. We gave a general argument that the metric on ${\cal M}$ is K\"ahler. It would be interesting to put further constraints on the general structure of the quantum corrected Zamolodchikov metric at finite 't Hooft coupling, which may guide us toward finding the stringy AdS dual of such SCFTs.

\subsection*{Acknowledgments}

We would like to thank F. Denef, D. Jafferis, S. Penati, and especially D. Gaiotto for discussions and correspondences.
X.Y. would like to thank S. Giombi for collaborations in unpublished work on closely related matters in CSM theories.
This work is supported in part by the Fundamental Laws Initiative Fund at Harvard University. 
X.Y. is supported by NSF Award PHY-0847457.

\appendix

\section{4-loop corrections to the Yukawa coupling}

In this appendix, we will explicitly compute the coefficient $c_1$ in (\ref{effective_superpotential}). The relevant supergraphs are listed in (\ref{ABC}). In practice, we find it easier to work with ordinary Feynman diagrams. In component fields, the potential 4-loop contribution to the 1PI effective superpotential is written as
\ie
&c_{1}M\frac{4\pi^2}{k^2}N^4\alpha_{ijmn}\bar\alpha^{mnpq}\alpha_{pqkl}\int{\rm Tr}\left(\Phi^i\Phi^j\Phi^k\Phi^l\right)d^2\theta\\
&=c_{1}M\frac{4\pi^2}{k^2}N^4\alpha_{ijmn}\bar\alpha^{mnpq}\alpha_{pqkl}\left[2{\rm Tr}\left(F^i\phi^j\phi^k\phi^l\right)+2{\rm Tr}\left(\phi^i\phi^j\phi^k F^l\right)\right.
\\&\left.-2{\rm Tr}\left(\psi^i\psi^j\phi^k\phi^l\right)-2{\rm Tr}\left(\psi^i\phi^j\phi^k\psi^l\right)-{\rm Tr}\left(\psi^i\phi^j\psi^k\phi^l\right)-{\rm Tr}\left(\phi^i\psi^j\phi^k\psi^l\right)\right]
\label{potential}
\fe
Since the various Yukawa coupling terms in the last line of (\ref{potential}) are related by supersymmetry, it is sufficient to compute one of them. Nevertheless, we will also include here the computations of the other terms as a cross check. We will then group the diagrams that contribute to the Yukawa coupling according to the supergraphs (\ref{ABC}). We will find that the contributions from the three types of supergraphs sum up to zero.

The supervertices appearing in (\ref{ABC}) correspond to the following interaction vertices of component fields 

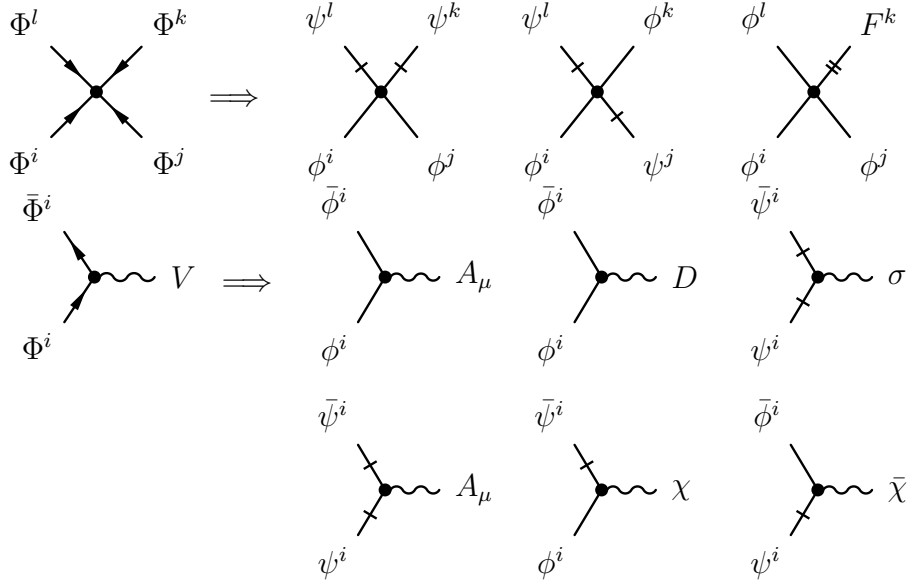
\bea\nonumber
\parbox{15mm}{\begin{fmffile}{S1}
        \begin{tabular}{c}
            \begin{fmfgraph*}(15,15)
                \fmfstraight
                \fmfset{arrow_len}{.3cm}\fmfset{arrow_ang}{12}
                \fmfleft{i1,i2}
                \fmfright{o1,o2}
                \fmflabel{$\Phi^i$}{i1}
                \fmflabel{$\Phi^l$}{i2}
                \fmflabel{$\Phi^j$}{o1}
                \fmflabel{$\Phi^k$}{o2}
                \fmf{fermion}{i1,d}
                \fmf{fermion}{i2,d}
                \fmf{fermion}{o1,d}
                \fmf{fermion}{o2,d}
                \fmfdot{d}
            \end{fmfgraph*}
        \end{tabular}
       \end{fmffile}}~~~~~
\Longrightarrow~~~~~
\parbox{15mm}{\begin{fmffile}{F1}
        \begin{tabular}{c}
            \begin{fmfgraph*}(15,15)
       \fmfcmd{
   vardef bar (expr p, len, ang) =
    ((-len/2,0)--(len/2,0))
       rotated (ang + angle
         direction length(p)/2 of p)
       shifted point length(p)/2 of p
   enddef;
   style_def tfermion expr p =
    draw p;
   ccutdraw bar (p, 2mm,  90)
  enddef;}
                \fmfleft{i1,i2}
                \fmfright{o1,o2}
                \fmflabel{$\phi^i$}{i1}
                \fmflabel{$\psi^l$}{i2}
                \fmflabel{$\phi^j$}{o1}
                \fmflabel{$\psi^k$}{o2}
                \fmf{plain}{i1,d}
                \fmf{tfermion}{i2,d}
                \fmf{plain}{o1,d}
                \fmf{tfermion}{o2,d}
                \fmfdot{d}
            \end{fmfgraph*}
        \end{tabular}
        \end{fmffile}}~~~~~~~~~~
\parbox{15mm}{\begin{fmffile}{F2}
        \begin{tabular}{c}
            \begin{fmfgraph*}(15,15)
       \fmfcmd{
   vardef bar (expr p, len, ang) =
    ((-len/2,0)--(len/2,0))
       rotated (ang + angle
         direction length(p)/2 of p)
       shifted point length(p)/2 of p
   enddef;
   style_def tfermion expr p =
    draw p;
   ccutdraw bar (p, 2mm,  90)
  enddef;}
                \fmfleft{i1,i2}
                \fmfright{o1,o2}
                \fmflabel{$\phi^i$}{i1}
                \fmflabel{$\psi^l$}{i2}
                \fmflabel{$\psi^j$}{o1}
                \fmflabel{$\phi^k$}{o2}
                \fmf{plain}{i1,d}
                \fmf{tfermion}{i2,d}
                \fmf{tfermion}{o1,d}
                \fmf{plain}{o2,d}
                \fmfdot{d}
            \end{fmfgraph*}
        \end{tabular}
        \end{fmffile}}~~~~~~~~~~
\parbox{15mm}{\begin{fmffile}{F3}
        \begin{tabular}{c}
            \begin{fmfgraph*}(15,15)
           \fmfcmd{
   vardef fbar (expr p, len, ang) =
    ((-len/2,0)--(len/2,0))
       rotated (ang + angle
         direction length(p)/2 of p)
       shifted point length(p)*45/100 of p
   enddef;
   vardef bbar (expr p, len, ang) =
    ((-len/2,0)--(len/2,0))
       rotated (ang + angle
         direction length(p)/2 of p)
       shifted point length(p)*55/100 of p
   enddef;
   style_def ttfermion expr p =
    draw p;
   ccutdraw fbar (p, 2mm,  90);
   ccutdraw bbar (p, 2mm,  90)
  enddef;}
                \fmfleft{i1,i2}
                \fmfright{o1,o2}
                \fmflabel{$\phi^i$}{i1}
                \fmflabel{$\phi^l$}{i2}
                \fmflabel{$\phi^j$}{o1}
                \fmflabel{$F^k$}{o2}
                \fmf{plain}{i1,d}
                \fmf{plain}{i2,d}
                \fmf{plain}{o1,d}
                \fmf{ttfermion}{o2,d}
                \fmfdot{d}
            \end{fmfgraph*}
        \end{tabular}
        \end{fmffile}}
\label{Fvertices}
\eea

\bigskip

\ie\nonumber
\parbox{15mm}{\begin{fmffile}{S2}
        \begin{tabular}{c}
            \begin{fmfgraph*}(15,15)
                \fmfstraight
                \fmfset{arrow_len}{.3cm}\fmfset{arrow_ang}{12}
                \fmfleft{i1,i2}
                \fmfright{o}
                \fmflabel{$\Phi^i$}{i1}
                \fmflabel{$\bar\Phi^i$}{i2}
                \fmflabel{$V$}{o}
                \fmf{fermion}{i1,d,i2}
                \fmf{photon}{o,d}
                \fmfdot{d}
            \end{fmfgraph*}
        \end{tabular}
       \end{fmffile}}~~~~~
\Longrightarrow~~~~~
&\parbox{15mm}{\begin{fmffile}{D1}
        \begin{tabular}{c}
            \begin{fmfgraph*}(15,15)
                \fmfleft{i1,i2}
                \fmfright{o}
                \fmflabel{$\phi^i$}{i1}
                \fmflabel{$\bar\phi^i$}{i2}
                \fmflabel{$A_\m$}{o}
                \fmf{plain}{i1,d,i2}
                \fmf{photon}{o,d}
                \fmfdot{d}
            \end{fmfgraph*}
        \end{tabular}
       \end{fmffile}}~~~~~~~~~~
\parbox{15mm}{\begin{fmffile}{D2}
        \begin{tabular}{c}
            \begin{fmfgraph*}(15,15)
                \fmfleft{i1,i2}
                \fmfright{o}
                \fmflabel{$\phi^i$}{i1}
                \fmflabel{$\bar\phi^i$}{i2}
                \fmflabel{$D$}{o}
                \fmf{plain}{i1,d,i2}
                \fmf{photon}{o,d}
                \fmfdot{d}
            \end{fmfgraph*}
        \end{tabular}
       \end{fmffile}}~~~~~~~~~~
\parbox{15mm}{\begin{fmffile}{D3}
        \begin{tabular}{c}
            \begin{fmfgraph*}(15,15)
       \fmfcmd{
   vardef bar (expr p, len, ang) =
    ((-len/2,0)--(len/2,0))
       rotated (ang + angle
         direction length(p)/2 of p)
       shifted point length(p)/2 of p
   enddef;
   style_def tfermion expr p =
    draw p;
   ccutdraw bar (p, 2mm,  90)
  enddef;}
                \fmfleft{i1,i2}
                \fmfright{o}
                \fmflabel{$\psi^i$}{i1}
                \fmflabel{$\bar\psi^i$}{i2}
                \fmflabel{$\sigma$}{o}
                \fmf{tfermion}{i1,d,i2}
                \fmf{photon}{o,d}
                \fmfdot{d}
            \end{fmfgraph*}
        \end{tabular}
       \end{fmffile}}\\\\\\
&\parbox{15mm}{\begin{fmffile}{D4}
        \begin{tabular}{c}
            \begin{fmfgraph*}(15,15)
       \fmfcmd{
   vardef bar (expr p, len, ang) =
    ((-len/2,0)--(len/2,0))
       rotated (ang + angle
         direction length(p)/2 of p)
       shifted point length(p)/2 of p
   enddef;
   style_def tfermion expr p =
    draw p;
   ccutdraw bar (p, 2mm,  90)
  enddef;}
                \fmfleft{i1,i2}
                \fmfright{o}
                \fmflabel{$\psi^i$}{i1}
                \fmflabel{$\bar\psi^i$}{i2}
                \fmflabel{$A_\m$}{o}
                \fmf{tfermion}{i1,d,i2}
                \fmf{photon}{o,d}
                \fmfdot{d}
            \end{fmfgraph*}
        \end{tabular}
       \end{fmffile}}~~~~~~~~~~
\parbox{15mm}{\begin{fmffile}{D5}
        \begin{tabular}{c}
            \begin{fmfgraph*}(15,15)
       \fmfcmd{
   vardef bar (expr p, len, ang) =
    ((-len/2,0)--(len/2,0))
       rotated (ang + angle
         direction length(p)/2 of p)
       shifted point length(p)/2 of p
   enddef;
   style_def tfermion expr p =
    draw p;
   ccutdraw bar (p, 2mm,  90)
  enddef;}
                \fmfleft{i1,i2}
                \fmfright{o}
                \fmflabel{$\phi^i$}{i1}
                \fmflabel{$\bar\psi^i$}{i2}
                \fmflabel{$\chi$}{o}
                \fmf{tfermion}{d,i2}
                \fmf{plain}{i1,d}
                \fmf{photon}{o,d}
                \fmfdot{d}
            \end{fmfgraph*}
        \end{tabular}
       \end{fmffile}}~~~~~~~~~~
\parbox{15mm}{\begin{fmffile}{D6}
        \begin{tabular}{c}
            \begin{fmfgraph*}(15,15)
       \fmfcmd{
   vardef bar (expr p, len, ang) =
    ((-len/2,0)--(len/2,0))
       rotated (ang + angle
         direction length(p)/2 of p)
       shifted point length(p)/2 of p
   enddef;
   style_def tfermion expr p =
    draw p;
   ccutdraw bar (p, 2mm,  90)
  enddef;}
                \fmfleft{i1,i2}
                \fmfright{o}
                \fmflabel{$\psi^i$}{i1}
                \fmflabel{$\bar\phi^i$}{i2}
                \fmflabel{$\bar\chi$}{o}
                \fmf{tfermion}{i1,d}
                \fmf{plain}{d,i2}
                \fmf{photon}{o,d}
                \fmfdot{d}
            \end{fmfgraph*}
        \end{tabular}
       \end{fmffile}}
\fe
\bigskip

\noindent
where the slashed line stands for the fermion $\psi^i$, and the double slashed line stands for the auxiliary field $F^i$. The wavy lines stand for a component field of the vector superfield. 
Next, we will make use of the one-loop corrected propagators for the vector gauge field and the auxiliary fields $D,\sigma,\chi$ by integrating out the matter fields. They are of the form
\bea\nonumber
\begin{aligned}
&\parbox{27mm}{\begin{fmffile}{lcp}
        \begin{tabular}{c}
            \begin{fmfgraph*}(27,15)
                \fmfstraight
                \fmfset{arrow_len}{.3cm}\fmfset{arrow_ang}{12}
                \fmfleft{i}
                \fmfright{o}
                \fmflabel{$A_\m$}{i}
                \fmflabel{$A_\m$}{o}
                \fmf{photon}{i,v1}
                \fmf{photon}{o,v1}
                \fmfblob{.2w}{v1}
            \end{fmfgraph*}
        \end{tabular}
\end{fmffile}}~~~~~ = ~~~~~
\parbox{27mm}{\begin{fmffile}{lcp1}
        \begin{tabular}{c}
            \begin{fmfgraph*}(27,15)
                \fmfleft{i}
                \fmfright{o}
                \fmflabel{$A_\m$}{i}
                \fmflabel{$A_\m$}{o}
                \fmf{photon}{i,d1}
                \fmf{photon}{o,d2}
                \fmf{plain,right=.5,tension=0.3}{d1,d2,d1}
            \end{fmfgraph*}
        \end{tabular}
\end{fmffile}}~~~~~ + ~~~~~
\parbox{27mm}{\begin{fmffile}{lcp2}
        \begin{tabular}{c}
            \begin{fmfgraph*}(27,15)
       \fmfcmd{
   vardef bar (expr p, len, ang) =
    ((-len/2,0)--(len/2,0))
       rotated (ang + angle
         direction length(p)/2 of p)
       shifted point length(p)/2 of p
   enddef;
   style_def tfermion expr p =
    draw p;
   ccutdraw bar (p, 2mm,  90)
  enddef;}
                \fmfleft{i}
                \fmfright{o}
                \fmflabel{$A_\m$}{i}
                \fmflabel{$A_\m$}{o}
                \fmf{photon}{i,d1}
                \fmf{photon}{o,d2}
                \fmf{tfermion,right=.5,tension=0.3}{d1,d2,d1}
            \end{fmfgraph*}
        \end{tabular}
\end{fmffile}}
\\ &
\parbox{27mm}{\begin{fmffile}{lcaf1}
        \begin{tabular}{c}
            \begin{fmfgraph*}(27,15)
                \fmfleft{i}
                \fmfright{o}
                \fmf{photon}{i,d1}
                \fmf{photon}{o,d2}
                \fmflabel{$D$}{i}
                \fmflabel{$D$}{o}
                \fmf{plain,right=.5,tension=0.3}{d1,d2,d1}
            \end{fmfgraph*}
        \end{tabular}
\end{fmffile}}
~~~~~~~~~~
\parbox{27mm}{\begin{fmffile}{lcaf3}
        \begin{tabular}{c}
            \begin{fmfgraph*}(27,15)
       \fmfcmd{
   vardef bar (expr p, len, ang) =
    ((-len/2,0)--(len/2,0))
       rotated (ang + angle
         direction length(p)/2 of p)
       shifted point length(p)/2 of p
   enddef;
   style_def tfermion expr p =
    draw p;
   ccutdraw bar (p, 2mm,  90)
  enddef;}
                \fmfleft{i}
                \fmfright{o}
                \fmflabel{$\chi$}{i}
                \fmflabel{$\bar\chi$}{o}
                \fmf{photon}{i,d1}
                \fmf{photon}{o,d2}
                \fmf{tfermion,right=.5,tension=0.3}{d1,d2}
                \fmf{plain,right=.5,tension=0.3}{d2,d1}
            \end{fmfgraph*}
        \end{tabular}
\end{fmffile}}
~~~~~~~~~~
\parbox{27mm}{\begin{fmffile}{lcaf2}
        \begin{tabular}{c}
            \begin{fmfgraph*}(27,15)
       \fmfcmd{
   vardef bar (expr p, len, ang) =
    ((-len/2,0)--(len/2,0))
       rotated (ang + angle
         direction length(p)/2 of p)
       shifted point length(p)/2 of p
   enddef;
   style_def tfermion expr p =
    draw p;
   ccutdraw bar (p, 2mm,  90)
  enddef;}
                \fmfleft{i}
                \fmfright{o}
                \fmf{photon}{i,d1}
                \fmf{photon}{o,d2}
                \fmflabel{$\sigma$}{i}
                \fmflabel{$\sigma$}{o}
                \fmf{tfermion,right=.5,tension=0.3}{d1,d2,d1}
            \end{fmfgraph*}
        \end{tabular}
\end{fmffile}} 
\end{aligned}
\eea
After integrating out the auxiliary fields, the loop-corrected auxiliary field propagators are replaced by a bubble, involving either the scalar, fermion, or both. They will be denoted by a dashed line or a slashed dashed line.
\bea\nonumber
\parbox{24mm}{\begin{fmffile}{dashes}
        \begin{tabular}{c}
            \begin{fmfgraph*}(27,15)
                \fmfleft{i}
                \fmfright{o}
                \fmf{dashes}{i,o}
            \end{fmfgraph*}
        \end{tabular}
        \end{fmffile}}
 = 
\parbox{18mm}{\begin{fmffile}{lcaf1a}
        \begin{tabular}{c}
            \begin{fmfgraph*}(18,6)
                \fmfleft{i}
                \fmfright{o}
                \fmf{plain,right=.5,tension=0.3}{i,o,i}
            \end{fmfgraph*}
        \end{tabular}
\end{fmffile}}{\rm~~ or~~}
\parbox{18mm}{\begin{fmffile}{lcaf2a}
        \begin{tabular}{c}
            \begin{fmfgraph*}(18,6)
       \fmfcmd{
   vardef bar (expr p, len, ang) =
    ((-len/2,0)--(len/2,0))
       rotated (ang + angle
         direction length(p)/2 of p)
       shifted point length(p)/2 of p
   enddef;
   style_def tfermion expr p =
    draw p;
   ccutdraw bar (p, 2mm,  90)
  enddef;}
                \fmfleft{i}
                \fmfright{o}
                \fmf{tfermion,right=.5,tension=0.3}{i,o,i}
            \end{fmfgraph*}
        \end{tabular}
\end{fmffile}}
~~~~~{\rm and}~~~~~
\parbox{24mm}{\begin{fmffile}{dashfermion}
        \begin{tabular}{c}
            \begin{fmfgraph*}(27,15)
           \fmfcmd{
   vardef bar (expr p, len, ang) =
    ((-len/2,0)--(len/2,0))
       rotated (ang + angle
         direction length(p)/2 of p)
       shifted point length(p)/2 of p
   enddef;
   style_def dashfermion expr p =
    draw_dashes p;
   ccutdraw bar (p, 2mm,  90)
  enddef;}
                \fmfleft{i}
                \fmfright{o}
                \fmf{dashfermion}{i,o}
            \end{fmfgraph*}
        \end{tabular}
        \end{fmffile}}
=
\parbox{18mm}{\begin{fmffile}{lcaf3a}
        \begin{tabular}{c}
            \begin{fmfgraph*}(18,6)
       \fmfcmd{
   vardef bar (expr p, len, ang) =
    ((-len/2,0)--(len/2,0))
       rotated (ang + angle
         direction length(p)/2 of p)
       shifted point length(p)/2 of p
   enddef;
   style_def tfermion expr p =
    draw p;
   ccutdraw bar (p, 2mm,  90)
  enddef;}
                \fmfleft{i}
                \fmfright{o}
                \fmf{tfermion,right=.5,tension=0.3}{i,o}
                \fmf{plain,right=.5,tension=0.3}{o,i}
            \end{fmfgraph*}
        \end{tabular}
\end{fmffile}}
\eea
Note that if the dashed line connects a pair of fermion lines, the bubble involves scalars, whereas if the dashed line connects a pair of scalar line, the bubble involves fermions only.

The relevant Feynman diagrams either involves a matter-loop-corrected auxiliary field propagator (represented by a dashed or slash-dashed line), or a matter-loop-corrected gauge field propagator. The former type of diagrams can be easily evaluated using the graphical rules described in the next section. To illustrate this, let us consider an example:
\be
\begin{aligned}
\parbox{45mm}{\begin{fmffile}{ex5}
        \begin{tabular}{c}
            \begin{fmfgraph*}(45,25)
       \fmfcmd{
   vardef bar (expr p, len, ang) =
    ((-len/2,0)--(len/2,0))
       rotated (ang + angle
         direction length(p)/2 of p)
       shifted point length(p)/2 of p
   enddef;
   style_def tfermion expr p =
    draw p;
   ccutdraw bar (p, 2mm,  90)
  enddef;}
                \fmfleft{i1,i2}
                \fmfright{o1,o2}
                \fmf{plain}{i2,v1}
                \fmf{tfermion}{i1,v1}
                \fmf{plain}{v2,o2}
                \fmf{tfermion}{v2,o1}
                \fmffixed{(.6h,0)}{v1,v3}
                \fmffixed{(.6h,0)}{v3,v2}
                \fmffixed{(-.3h,.25h)}{v3,v4}
                \fmffixed{(.3h,.25h)}{v3,v5}
                \fmf{tfermion,right=.7,tension=.3}{v1,v3}
                \fmf{tfermion,right=.7,tension=.3}{v3,v2}
                \fmf{plain,left=.3,tension=.3}{v1,v4,v3}
                \fmf{plain,left=.3,tension=.3}{v3,v5,v2}
                \fmf{dashes,left=.7,tension=0}{v4,v5}
             \end{fmfgraph*}
        \end{tabular}
        \end{fmffile}} & =
\parbox{45mm}{\begin{fmffile}{ex5a}
        \begin{tabular}{c}
            \begin{fmfgraph*}(45,35)
       \fmfcmd{
   vardef bar (expr p, len, ang) =
    ((-len/2,0)--(len/2,0))
       rotated (ang + angle
         direction length(p)/2 of p)
       shifted point length(p)/2 of p
   enddef;
   style_def tfermion expr p =
    draw p;
   ccutdraw bar (p, 2mm,  90)
  enddef;}
                \fmfleft{v1}
                \fmfright{v2}
                %\fmffixed{(.6h,0)}{v1,v3}
                %\fmffixed{(.6h,0)}{v3,v2}
                \fmffixed{(-.25h,.3h)}{v3,v4}
                \fmffixed{(.25h,.3h)}{v3,v5}
                \fmf{tfermion,right=.7,tension=.3,label=2}{v1,v3}
                \fmf{tfermion,right=.7,tension=.3,label=2}{v3,v2}
                \fmf{plain,left=.3,tension=.3,label=2,label.side=right,label.dist=2}{v1,v4,v3}
                \fmf{plain,left=.3,tension=.3,label=2, label.side=right,label.dist=2}{v3,v5,v2}
                \fmf{plain,left=.7,tension=0,label=$2-d$}{v4,v5}
                \fmfdot{v1,v2,v3,v4,v5}
             \end{fmfgraph*}
        \end{tabular}
        \end{fmffile}}A_{ff}(2,2)\\
& = \parbox{35mm}{\begin{fmffile}{ex5b}
        \begin{tabular}{c}
            \begin{fmfgraph*}(35,35)
       \fmfcmd{
   vardef bar (expr p, len, ang) =
    ((-len/2,0)--(len/2,0))
       rotated (ang + angle
         direction length(p)/2 of p)
       shifted point length(p)/2 of p
   enddef;
   style_def tfermion expr p =
    draw p;
   ccutdraw bar (p, 2mm,  90)
  enddef;}
                \fmfleft{v1}
                \fmfright{v2}
                \fmfbottom{v3}
                \fmf{tfermion,right=.3,tension=.3,label=4}{v1,v3}
                \fmf{tfermion,right=.3,tension=.3,label=4}{v3,v2}
                \fmf{plain,left=.3,tension=.3,label=2,label.side=left,label.dist=2}{v1,v3}
                \fmf{plain,left=.3,tension=.3,label=2, label.side=left,label.dist=2}{v3,v2}
                \fmf{plain,left=.3,tension=0,label=$2-d$}{v1,v2}
                \fmfdot{v1,v2,v3}
             \end{fmfgraph*}
        \end{tabular}
        \end{fmffile}}A_{ff}(2,2)\\ & = A_{ff}(2,2)A_{fs}(4,2)^2 (-1)\times
\parbox{35mm}{\begin{fmffile}{ex5c}
        \begin{tabular}{c}
            \begin{fmfgraph*}(35,35)
                \fmftop{v1}
                \fmfbottom{v2}
                \fmflabel{$12-3d$}{v4}
                \fmf{phantom}{v1,v3}
                \fmf{phantom}{v2,v4}
                \fmf{plain,tension=.3,left=1}{v3,v4}
                \fmf{plain,tension=.3,left=1}{v4,v3}
                \fmfdot{v3}
             \end{fmfgraph*}
        \end{tabular}
        \end{fmffile}}
\end{aligned}
\ee
where $A_{ss}, A_{fs}$ and $A_{ff}$ are coefficients that will be defined in the next section. The remaining scalar loop gives a standard logarithmic divergence at $d=3$,
\bea
\parbox{25mm}{\begin{fmffile}{ex5c}
        \begin{tabular}{c}
            \begin{fmfgraph*}(35,35)
                \fmftop{v1}
                \fmfbottom{v2}
                \fmflabel{$12-3d$}{v4}
                \fmf{phantom}{v1,v3}
                \fmf{phantom}{v2,v4}
                \fmf{plain,tension=.3,left=1}{v3,v4}
                \fmf{plain,tension=.3,left=1}{v4,v3}
                \fmfdot{v3}
             \end{fmfgraph*}
        \end{tabular}
        \end{fmffile}} \to  {\log\Lambda\over 2\pi^2}
\label{common_loop}
\eea
(\ref{common_loop}) is a common factor for all graphs we will encounter, and will thus be omitted in the following. 

The results for the diagrams that involve a matter-loop-corrected auxiliary field are listed as follows. 
\bea
A_ 1 = \parbox{45mm}{\begin{fmffile}{A1}
        \begin{tabular}{c}
            \begin{fmfgraph*}(45,25)
       \fmfcmd{
   vardef bar (expr p, len, ang) =
    ((-len/2,0)--(len/2,0))
       rotated (ang + angle
         direction length(p)/2 of p)
       shifted point length(p)/2 of p
   enddef;
   style_def tfermion expr p =
    draw p;
   ccutdraw bar (p, 2mm,  90)
  enddef;}
                \fmfleft{i1,i2}
                \fmfright{o1,o2}
                \fmf{plain}{i2,v1}
                \fmf{tfermion}{i1,v1}
                \fmf{plain}{v2,o2}
                \fmf{tfermion}{v2,o1}
                \fmffixed{(.6h,0)}{v1,v3}
                \fmffixed{(.6h,0)}{v3,v2}
                \fmffixed{(-.3h,.25h)}{v3,v4}
                \fmffixed{(.3h,.25h)}{v3,v5}
                \fmf{tfermion,right=.7,tension=.3}{v1,v3}
                \fmf{tfermion,right=.7,tension=.3}{v3,v2}
                \fmf{plain,left=.3,tension=.3}{v1,v4,v3}
                \fmf{plain,left=.3,tension=.3}{v3,v5,v2}
                \fmf{dashes,left=.7,tension=0}{v4,v5}
             \end{fmfgraph*}
        \end{tabular}
        \end{fmffile}} & = & -{A_{ff}}(2,2){A_{fs}}(2,2)^2 \\
A_2 = \parbox{45mm}{\begin{fmffile}{A2}
        \begin{tabular}{c}
            \begin{fmfgraph*}(45,25)
       \fmfcmd{
   vardef bar (expr p, len, ang) =
    ((-len/2,0)--(len/2,0))
       rotated (ang + angle
         direction length(p)/2 of p)
       shifted point length(p)/2 of p
   enddef;
   style_def tfermion expr p =
    draw p;
   ccutdraw bar (p, 2mm,  90)
  enddef;}
         \fmfcmd{
   vardef bar (expr p, len, ang) =
    ((-len/2,0)--(len/2,0))
       rotated (ang + angle
         direction length(p)/2 of p)
       shifted point length(p)/2 of p
   enddef;
   style_def dashfermion expr p =
    draw_dashes p;
   ccutdraw bar (p, 2mm,  90)
  enddef;}
                \fmfleft{i1,i2}
                \fmfright{o1,o2}
                \fmf{tfermion}{i2,v1}
                \fmf{tfermion}{i1,v1}
                \fmf{plain}{v2,o2}
                \fmf{plain}{v2,o1}
                \fmffixed{(.6h,0)}{v1,v3}
                \fmffixed{(.6h,0)}{v3,v2}
                \fmffixed{(-.3h,.25h)}{v3,v4}
                \fmffixed{(.3h,.25h)}{v3,v5}
                \fmf{plain,right=.7,tension=.3}{v1,v3}
                \fmf{tfermion,right=.7,tension=.3}{v3,v2}
                \fmf{plain,left=.3,tension=.3}{v1,v4}
                \fmf{tfermion,left=.3,tension=.3}{v4,v3}
                \fmf{plain,left=.3,tension=.3}{v3,v5}
                \fmf{tfermion,left=.3,tension=.3}{v5,v2}
                \fmf{dashfermion,left=.7,tension=0}{v4,v5}
             \end{fmfgraph*}
        \end{tabular}
        \end{fmffile}} & = & -2{A_{fs}}(2,2){A_{fs}}(2,4){A_{ss}}(2,2)\\
A_3 = \parbox{45mm}{\begin{fmffile}{A3}
        \begin{tabular}{c}
            \begin{fmfgraph*}(45,25)
       \fmfcmd{
   vardef bar (expr p, len, ang) =
    ((-len/2,0)--(len/2,0))
       rotated (ang + angle
         direction length(p)/2 of p)
       shifted point length(p)/2 of p
   enddef;
   style_def tfermion expr p =
    draw p;
   ccutdraw bar (p, 2mm,  90)
  enddef;}
         \fmfcmd{
   vardef bar (expr p, len, ang) =
    ((-len/2,0)--(len/2,0))
       rotated (ang + angle
         direction length(p)/2 of p)
       shifted point length(p)/2 of p
   enddef;
   style_def dashfermion expr p =
    draw_dashes p;
   ccutdraw bar (p, 2mm,  90)
  enddef;}
                \fmfleft{i1,i2}
                \fmfright{o1,o2}
                \fmf{plain}{i2,v1}
                \fmf{tfermion}{i1,v1}
                \fmf{plain}{v2,o2}
                \fmf{tfermion}{v2,o1}
                \fmffixed{(.6h,0)}{v1,v3}
                \fmffixed{(.6h,0)}{v3,v2}
                \fmffixed{(-.3h,.25h)}{v3,v4}
                \fmffixed{(.3h,.25h)}{v3,v5}
                \fmf{tfermion,right=.7,tension=.3}{v1,v3}
                \fmf{plain,right=.7,tension=.3}{v3,v2}
                \fmf{plain,left=.3,tension=.3}{v1,v4}
                \fmf{tfermion,left=.3,tension=.3}{v4,v3}
                \fmf{plain,left=.3,tension=.3}{v3,v5}
                \fmf{tfermion,left=.3,tension=.3}{v5,v2}
                \fmf{dashfermion,left=.7,tension=0}{v4,v5}
             \end{fmfgraph*}
        \end{tabular}
        \end{fmffile}} & = & \frac{1}{2}{A_{fs}}(2,2){A_{fs}}(4,2){A_{ff}}(4,2) \\
A_4 = \parbox{45mm}{\begin{fmffile}{A4}
        \begin{tabular}{c}
            \begin{fmfgraph*}(45,25)
       \fmfcmd{
   vardef bar (expr p, len, ang) =
    ((-len/2,0)--(len/2,0))
       rotated (ang + angle
         direction length(p)/2 of p)
       shifted point length(p)/2 of p
   enddef;
   style_def tfermion expr p =
    draw p;
   ccutdraw bar (p, 2mm,  90)
  enddef;}
                \fmfleft{i1,i2}
                \fmfright{o1,o2}
                \fmf{plain}{i2,v1}
                \fmf{tfermion}{i1,v1}
                \fmf{plain}{v2,o2}
                \fmf{tfermion}{v2,o1}
                \fmffixed{(.6h,0)}{v1,v3}
                \fmffixed{(.6h,0)}{v3,v2}
                \fmffixed{(-.3h,.25h)}{v3,v4}
                \fmffixed{(.3h,.25h)}{v3,v5}
                \fmf{plain,right=.7,tension=.3}{v1,v3}
                \fmf{plain,right=.7,tension=.3}{v3,v2}
                \fmf{tfermion,left=.3,tension=.3}{v1,v4,v3}
                \fmf{tfermion,left=.3,tension=.3}{v3,v5,v2}
                \fmf{dashes,left=.7,tension=0}{v4,v5}
             \end{fmfgraph*}
        \end{tabular}
        \end{fmffile}} & = & \frac{1}{4}{A_{ff}}(4,2)^2 {A_{ss}}(2,2)
\eea

\bea
B_1 = \parbox{45mm}{\begin{fmffile}{B1}
        \begin{tabular}{c}
            \begin{fmfgraph*}(45,25)
       \fmfcmd{
   vardef bar (expr p, len, ang) =
    ((-len/2,0)--(len/2,0))
       rotated (ang + angle
         direction length(p)/2 of p)
       shifted point length(p)/2 of p
   enddef;
   style_def tfermion expr p =
    draw p;
   ccutdraw bar (p, 2mm,  90)
  enddef;}
                \fmfleft{i1,i2}
                \fmfright{o1,o2}
                \fmf{plain}{i2,v4,v1}
                \fmf{tfermion}{i1,v1}
                \fmf{plain}{v2,o2}
                \fmf{tfermion}{v2,o1}
                \fmffixed{(.6h,0)}{v1,v3}
                \fmffixed{(.6h,0)}{v3,v2}
               %\fmffixed{(-.3h,.25h)}{v3,v4}
                \fmffixed{(.3h,.25h)}{v3,v5}
                \fmf{tfermion,right=.7,tension=.3}{v1,v3}
                \fmf{tfermion,right=.7,tension=.3}{v3,v2}
                \fmf{plain,left=.7,tension=.3}{v1,v3}
                \fmf{plain,left=.3,tension=.3}{v3,v5,v2}
                \fmf{dashes,left=.35,tension=0}{v4,v5}
             \end{fmfgraph*}
        \end{tabular}
        \end{fmffile}} & = & -{A_{ff}}(2,2){A_{fs}}(2,2){A_{fs}}(4,2) \\
B_2 = \parbox{45mm}{\begin{fmffile}{B2}
        \begin{tabular}{c}
            \begin{fmfgraph*}(45,25)
       \fmfcmd{
   vardef bar (expr p, len, ang) =
    ((-len/2,0)--(len/2,0))
       rotated (ang + angle
         direction length(p)/2 of p)
       shifted point length(p)/2 of p
   enddef;
   style_def tfermion expr p =
    draw p;
   ccutdraw bar (p, 2mm,  90)
  enddef;}
           \fmfcmd{
   vardef bar (expr p, len, ang) =
    ((-len/2,0)--(len/2,0))
       rotated (ang + angle
         direction length(p)/2 of p)
       shifted point length(p)/2 of p
   enddef;
   style_def dashfermion expr p =
    draw_dashes p;
   ccutdraw bar (p, 2mm,  90)
  enddef;}
                \fmfleft{i1,i2}
                \fmfright{o1,o2}
                \fmf{plain}{i2,v4}
                \fmf{tfermion}{v4,v1}
                \fmf{tfermion}{i1,v1}
                \fmf{plain}{v2,o2}
                \fmf{tfermion}{v2,o1}
                \fmffixed{(.6h,0)}{v1,v3}
                \fmffixed{(.6h,0)}{v3,v2}
               %\fmffixed{(-.3h,.25h)}{v3,v4}
                \fmffixed{(.3h,.25h)}{v3,v5}
                \fmf{plain,right=.7,tension=.3}{v1,v3}
                \fmf{tfermion,right=.7,tension=.3}{v3,v2}
                \fmf{plain,left=.7,tension=.3}{v1,v3}
                \fmf{tfermion,left=.3,tension=.3}{v3,v5}
                \fmf{plain,left=.3,tension=.3}{v5,v2}
                \fmf{dashfermion,left=.35,tension=0}{v4,v5}
             \end{fmfgraph*}
        \end{tabular}
        \end{fmffile}} & = & \frac{1}{2}{A_{fs}}(2,2){A_{ss}}(2,2){A_{ff}}(4,2)\\
B_3 = \parbox{45mm}{\begin{fmffile}{B3}
        \begin{tabular}{c}
            \begin{fmfgraph*}(45,25)
       \fmfcmd{
   vardef bar (expr p, len, ang) =
    ((-len/2,0)--(len/2,0))
       rotated (ang + angle
         direction length(p)/2 of p)
       shifted point length(p)/2 of p
   enddef;
   style_def tfermion expr p =
    draw p;
   ccutdraw bar (p, 2mm,  90)
  enddef;}
           \fmfcmd{
   vardef bar (expr p, len, ang) =
    ((-len/2,0)--(len/2,0))
       rotated (ang + angle
         direction length(p)/2 of p)
       shifted point length(p)/2 of p
   enddef;
   style_def dashfermion expr p =
    draw_dashes p;
   ccutdraw bar (p, 2mm,  90)
  enddef;}
                \fmfleft{i1,i2}
                \fmfright{o1,o2}
                \fmf{tfermion}{i2,v4}
                \fmf{plain}{v4,v1}
                \fmf{tfermion}{i1,v1}
                \fmf{plain}{v2,o2}
                \fmf{plain}{v2,o1}
                \fmffixed{(.6h,0)}{v1,v3}
                \fmffixed{(.6h,0)}{v3,v2}
               %\fmffixed{(-.3h,.25h)}{v3,v4}
                \fmffixed{(.3h,.25h)}{v3,v5}
                \fmf{tfermion,right=.7,tension=.3}{v1,v3}
                \fmf{tfermion,right=.7,tension=.3}{v3,v2}
                \fmf{plain,left=.7,tension=.3}{v1,v3}
                \fmf{plain,left=.3,tension=.3}{v3,v5}
                \fmf{tfermion,left=.3,tension=.3}{v5,v2}
                \fmf{dashfermion,left=.35,tension=0}{v4,v5}
             \end{fmfgraph*}
        \end{tabular}
        \end{fmffile}} & = & -{A_{ff}}(2,2){A_{fs}}(2,2){A_{fs}}(4,2)
\eea

\bea
B_4 = \parbox{45mm}{\begin{fmffile}{B4}
        \begin{tabular}{c}
            \begin{fmfgraph*}(45,25)
       \fmfcmd{
   vardef bar (expr p, len, ang) =
    ((-len/2,0)--(len/2,0))
       rotated (ang + angle
         direction length(p)/2 of p)
       shifted point length(p)/2 of p
   enddef;
   style_def tfermion expr p =
    draw p;
   ccutdraw bar (p, 2mm,  90)
  enddef;}
           \fmfcmd{
   vardef bar (expr p, len, ang) =
    ((-len/2,0)--(len/2,0))
       rotated (ang + angle
         direction length(p)/2 of p)
       shifted point length(p)/2 of p
   enddef;
   style_def dashfermion expr p =
    draw_dashes p;
   ccutdraw bar (p, 2mm,  90)
  enddef;}
                \fmfleft{i1,i2}
                \fmfright{o1,o2}
                \fmf{plain}{i2,v4}
                \fmf{tfermion}{v4,v1}
                \fmf{plain}{i1,v1}
                \fmf{tfermion}{v2,o2}
                \fmf{tfermion}{v2,o1}
                \fmffixed{(.6h,0)}{v1,v3}
                \fmffixed{(.6h,0)}{v3,v2}
               %\fmffixed{(-.3h,.25h)}{v3,v4}
                \fmffixed{(.3h,.25h)}{v3,v5}
                \fmf{plain,right=.7,tension=.3}{v1,v3}
                \fmf{plain,right=.7,tension=.3}{v3,v2}
                \fmf{tfermion,left=.7,tension=.3}{v1,v3}
                \fmf{tfermion,left=.3,tension=.3}{v3,v5}
                \fmf{plain,left=.3,tension=.3}{v5,v2}
                \fmf{dashfermion,left=.35,tension=0}{v4,v5}
             \end{fmfgraph*}
        \end{tabular}
        \end{fmffile}} & = & -2{A_{fs}}(2,2)^2 {A_{fs}}(2,4) \\
B_5 = \parbox{45mm}{\begin{fmffile}{B5}
        \begin{tabular}{c}
            \begin{fmfgraph*}(45,25)
       \fmfcmd{
   vardef bar (expr p, len, ang) =
    ((-len/2,0)--(len/2,0))
       rotated (ang + angle
         direction length(p)/2 of p)
       shifted point length(p)/2 of p
   enddef;
   style_def tfermion expr p =
    draw p;
   ccutdraw bar (p, 2mm,  90)
  enddef;}
           \fmfcmd{
   vardef bar (expr p, len, ang) =
    ((-len/2,0)--(len/2,0))
       rotated (ang + angle
         direction length(p)/2 of p)
       shifted point length(p)/2 of p
   enddef;
   style_def dashfermion expr p =
    draw_dashes p;
   ccutdraw bar (p, 2mm,  90)
  enddef;}
                \fmfleft{i1,i2}
                \fmfright{o1,o2}
                \fmf{tfermion}{i2,v4}
                \fmf{tfermion}{v4,v1}
                \fmf{plain}{i1,v1}
                \fmf{plain}{v2,o2}
                \fmf{tfermion}{v2,o1}
                \fmffixed{(.6h,0)}{v1,v3}
                \fmffixed{(.6h,0)}{v3,v2}
               %\fmffixed{(-.3h,.25h)}{v3,v4}
                \fmffixed{(.3h,.25h)}{v3,v5}
                \fmf{plain,right=.7,tension=.3}{v1,v3}
                \fmf{plain,right=.7,tension=.3}{v3,v2}
                \fmf{tfermion,left=.7,tension=.3}{v1,v3}
                \fmf{tfermion,left=.3,tension=.3}{v3,v5}
                \fmf{tfermion,left=.3,tension=.3}{v5,v2}
                \fmf{dashes,left=.35,tension=0}{v4,v5}
             \end{fmfgraph*}
        \end{tabular}
        \end{fmffile}} & = & \frac{1}{2}{A_{ss}}(2,2){A_{fs}}(2,2){A_{ff}}(4,2) \\
B_4 = \parbox{45mm}{\begin{fmffile}{B6}
        \begin{tabular}{c}
            \begin{fmfgraph*}(45,25)
       \fmfcmd{
   vardef bar (expr p, len, ang) =
    ((-len/2,0)--(len/2,0))
       rotated (ang + angle
         direction length(p)/2 of p)
       shifted point length(p)/2 of p
   enddef;
   style_def tfermion expr p =
    draw p;
   ccutdraw bar (p, 2mm,  90)
  enddef;}
           \fmfcmd{
   vardef bar (expr p, len, ang) =
    ((-len/2,0)--(len/2,0))
       rotated (ang + angle
         direction length(p)/2 of p)
       shifted point length(p)/2 of p
   enddef;
   style_def dashfermion expr p =
    draw_dashes p;
   ccutdraw bar (p, 2mm,  90)
  enddef;}
                \fmfleft{i1,i2}
                \fmfright{o1,o2}
                \fmf{tfermion}{i2,v4}
                \fmf{plain}{v4,v1}
                \fmf{plain}{i1,v1}
                \fmf{tfermion}{v2,o2}
                \fmf{plain}{v2,o1}
                \fmffixed{(.6h,0)}{v1,v3}
                \fmffixed{(.6h,0)}{v3,v2}
               %\fmffixed{(-.3h,.25h)}{v3,v4}
                \fmffixed{(.3h,.25h)}{v3,v5}
                \fmf{tfermion,right=.7,tension=.3}{v1,v3}
                \fmf{plain,right=.7,tension=.3}{v3,v2}
                \fmf{tfermion,left=.7,tension=.3}{v1,v3}
                \fmf{plain,left=.3,tension=.3}{v3,v5}
                \fmf{tfermion,left=.3,tension=.3}{v5,v2}
                \fmf{dashfermion,left=.35,tension=0}{v4,v5}
             \end{fmfgraph*}
        \end{tabular}
        \end{fmffile}} & = & -{A_{fs}}(2,2){A_{fs}}(4,2){A_{ff}}(2,2)
\eea

\bea
C_1 = \parbox{45mm}{\begin{fmffile}{C1}
        \begin{tabular}{c}
            \begin{fmfgraph*}(45,25)
       \fmfcmd{
   vardef bar (expr p, len, ang) =
    ((-len/2,0)--(len/2,0))
       rotated (ang + angle
         direction length(p)/2 of p)
       shifted point length(p)/2 of p
   enddef;
   style_def tfermion expr p =
    draw p;
   ccutdraw bar (p, 2mm,  90)
  enddef;}
           \fmfcmd{
   vardef bar (expr p, len, ang) =
    ((-len/2,0)--(len/2,0))
       rotated (ang + angle
         direction length(p)/2 of p)
       shifted point length(p)/2 of p
   enddef;
   style_def dashfermion expr p =
    draw_dashes p;
   ccutdraw bar (p, 2mm,  90)
  enddef;}
                \fmfleft{i1,i2}
                \fmfright{o1,o2}
                \fmf{plain}{i2,v4}
                \fmf{plain}{v4,v1}
                \fmf{tfermion}{i1,v1}
                \fmf{plain}{v2,v5}
                \fmf{plain}{v5,o2}
                \fmf{tfermion}{v2,o1}
                \fmffixed{(.6h,0)}{v1,v3}
                \fmffixed{(.6h,0)}{v3,v2}
               %\fmffixed{(-.3h,.25h)}{v3,v4}
                %\fmffixed{(.3h,.25h)}{v3,v5}
                \fmf{tfermion,right=.7,tension=.3}{v1,v3}
                \fmf{tfermion,right=.7,tension=.3}{v3,v2}
                \fmf{plain,left=.7,tension=.3}{v1,v3}
                \fmf{plain,left=.7,tension=.3}{v3,v2}
                \fmf{dashes,left=.2,tension=0}{v4,v5}
             \end{fmfgraph*}
        \end{tabular}
        \end{fmffile}} & = & -{A_{fs}}(2,2)^2{A_{ff}}(2,2)\\
C_2 = \parbox{45mm}{\begin{fmffile}{C2}
        \begin{tabular}{c}
            \begin{fmfgraph*}(45,25)
       \fmfcmd{
   vardef bar (expr p, len, ang) =
    ((-len/2,0)--(len/2,0))
       rotated (ang + angle
         direction length(p)/2 of p)
       shifted point length(p)/2 of p
   enddef;
   style_def tfermion expr p =
    draw p;
   ccutdraw bar (p, 2mm,  90)
  enddef;}
           \fmfcmd{
   vardef bar (expr p, len, ang) =
    ((-len/2,0)--(len/2,0))
       rotated (ang + angle
         direction length(p)/2 of p)
       shifted point length(p)/2 of p
   enddef;
   style_def dashfermion expr p =
    draw_dashes p;
   ccutdraw bar (p, 2mm,  90)
  enddef;}
                \fmfleft{i1,i2}
                \fmfright{o1,o2}
                \fmf{plain}{i2,v4}
                \fmf{tfermion}{v4,v1}
                \fmf{tfermion}{i1,v1}
                \fmf{plain}{v2,v5}
                \fmf{tfermion}{v5,o2}
                \fmf{plain}{v2,o1}
                \fmffixed{(.6h,0)}{v1,v3}
                \fmffixed{(.6h,0)}{v3,v2}
               %\fmffixed{(-.3h,.25h)}{v3,v4}
                %\fmffixed{(.3h,.25h)}{v3,v5}
                \fmf{plain,right=.7,tension=.3}{v1,v3}
                \fmf{tfermion,right=.7,tension=.3}{v3,v2}
                \fmf{plain,left=.7,tension=.3}{v1,v3}
                \fmf{tfermion,left=.7,tension=.3}{v3,v2}
                \fmf{dashfermion,left=.2,tension=0}{v4,v5}
             \end{fmfgraph*}
        \end{tabular}
        \end{fmffile}} & = & -{A_{fs}}(2,2){A_{ss}}(2,2){A_{ff}}(2,2) \\
C_3 = \parbox{45mm}{\begin{fmffile}{C3}
        \begin{tabular}{c}
            \begin{fmfgraph*}(45,25)
       \fmfcmd{
   vardef bar (expr p, len, ang) =
    ((-len/2,0)--(len/2,0))
       rotated (ang + angle
         direction length(p)/2 of p)
       shifted point length(p)/2 of p
   enddef;
   style_def tfermion expr p =
    draw p;
   ccutdraw bar (p, 2mm,  90)
  enddef;}
           \fmfcmd{
   vardef bar (expr p, len, ang) =
    ((-len/2,0)--(len/2,0))
       rotated (ang + angle
         direction length(p)/2 of p)
       shifted point length(p)/2 of p
   enddef;
   style_def dashfermion expr p =
    draw_dashes p;
   ccutdraw bar (p, 2mm,  90)
  enddef;}
                \fmfleft{i1,i2}
                \fmfright{o1,o2}
                \fmf{tfermion}{i2,v4,v1}
                \fmf{plain}{i1,v1}
                \fmf{tfermion}{v2,v5,o2}
                \fmf{plain}{v2,o1}
                \fmffixed{(.6h,0)}{v1,v3}
                \fmffixed{(.6h,0)}{v3,v2}
               %\fmffixed{(-.3h,.25h)}{v3,v4}
                %\fmffixed{(.3h,.25h)}{v3,v5}
                \fmf{tfermion,right=.7,tension=.3}{v1,v3}
                \fmf{tfermion,right=.7,tension=.3}{v3,v2}
                \fmf{plain,left=.7,tension=.3}{v1,v3}
                \fmf{plain,left=.7,tension=.3}{v3,v2}
                \fmf{dashes,left=.2,tension=0}{v4,v5}
             \end{fmfgraph*}
        \end{tabular}
        \end{fmffile}} & = & {A_{ss}}(2,2){A_{fs}}(2,2)^2 \\
C_4 = \parbox{45mm}{\begin{fmffile}{C4}
        \begin{tabular}{c}
            \begin{fmfgraph*}(45,25)
       \fmfcmd{
   vardef bar (expr p, len, ang) =
    ((-len/2,0)--(len/2,0))
       rotated (ang + angle
         direction length(p)/2 of p)
       shifted point length(p)/2 of p
   enddef;
   style_def tfermion expr p =
    draw p;
   ccutdraw bar (p, 2mm,  90)
  enddef;}
           \fmfcmd{
   vardef bar (expr p, len, ang) =
    ((-len/2,0)--(len/2,0))
       rotated (ang + angle
         direction length(p)/2 of p)
       shifted point length(p)/2 of p
   enddef;
   style_def dashfermion expr p =
    draw_dashes p;
   ccutdraw bar (p, 2mm,  90)
  enddef;}
                \fmfleft{i1,i2}
                \fmfright{o1,o2}
                \fmf{tfermion}{i2,v4}
                \fmf{plain}{v4,v1}
                \fmf{tfermion}{i1,v1}
                \fmf{tfermion}{v2,v5}
                \fmf{plain}{v5,o2}
                \fmf{plain}{v2,o1}
                \fmffixed{(.6h,0)}{v1,v3}
                \fmffixed{(.6h,0)}{v3,v2}
               %\fmffixed{(-.3h,.25h)}{v3,v4}
                %\fmffixed{(.3h,.25h)}{v3,v5}
                \fmf{tfermion,right=.7,tension=.3}{v1,v3}
                \fmf{tfermion,right=.7,tension=.3}{v3,v2}
                \fmf{plain,left=.7,tension=.3}{v1,v3}
                \fmf{plain,left=.7,tension=.3}{v3,v2}
                \fmf{dashfermion,left=.2,tension=0}{v4,v5}
             \end{fmfgraph*}
        \end{tabular}
        \end{fmffile}} & = & {A_{fs}}(2,2)^3
\eea

Our graphical rules do not apply directly to the diagrams with a loop-corrected gauge field propagators.
However, many of them are zero due to the following simplication.
(1) A Chern-Simons gauge field propagator attached to an external scalar line (of zero momentum) vanishes,

\bea
\parbox{25mm}{\begin{fmffile}{obs1}
        \begin{tabular}{c}
            \begin{fmfgraph*}(25,25)
                \fmfleft{i1}
                \fmfright{o1}
                \fmfbottom{b1}
                \fmf{photon,right=.4,tension=0,label=$\leftarrow k$,label.side=right}{o1,v1}
                \fmf{plain,label.side=right}{i1,v1}
                \fmf{plain,label=$k\searrow$,label.side=right}{v1,b1}
            \end{fmfgraph*}
        \end{tabular}
\end{fmffile}} & \propto & k^\m k^\n \epsilon_{\m\n\rho} = 0
\eea
and  (2) a Chern-Simons propagator attached to the following purely scalar or purely fermion loop gives zero.

\bea
\parbox{30mm}{\begin{fmffile}{obs2}
        \begin{tabular}{c}
            \begin{fmfgraph*}(30,12)
       \fmfcmd{
   vardef bar (expr p, len, ang) =
    ((-len/2,0)--(len/2,0))
       rotated (ang + angle
         direction length(p)/2 of p)
       shifted point length(p)/2 of p
   enddef;
   style_def tfermion expr p =
    draw p;
   ccutdraw bar (p, 2mm,  90)
  enddef;}
                \fmfleft{v1}
                \fmfright{o1,o2}
                \fmffixed{(-.2h,-.2h)}{v1,i1}
                \fmffixed{(-.2h,.2h)}{v1,i2}
                \fmffixed{(2h,0)}{v1,v2}
                \fmffixed{(1h,.4h)}{v1,v3}
                \fmffixed{(1h,1.2h)}{v1,v4}
                \fmf{plain,right=.2}{i1,v1,i2}
                \fmf{plain}{o1,v2,o2}
                \fmf{photon,tension=0,label=$\downarrow k$,label.side=left}{v4,v3}
                \fmf{tfermion,right=.5,label=$p\rightarrow$,label.side=right}{v1,v2}
                \fmf{tfermion,left=.25,label=$\stackrel{p+k}{\leftarrow}$,label.side=left}{v1,v3}
                \fmf{tfermion,right=.25}{v2,v3}
            \end{fmfgraph*}
        \end{tabular}
\end{fmffile}}~~ & \propto & \int d^3 p\left(p^\m+k^\m\right)k^\n\epsilon_{\m\n\rho}p^2 = 0
\eea

\bea
\parbox{30mm}{\begin{fmffile}{obs2a}
        \begin{tabular}{c}
            \begin{fmfgraph*}(30,12)
       \fmfcmd{
   vardef bar (expr p, len, ang) =
    ((-len/2,0)--(len/2,0))
       rotated (ang + angle
         direction length(p)/2 of p)
       shifted point length(p)/2 of p
   enddef;
   style_def tfermion expr p =
    draw p;
   ccutdraw bar (p, 2mm,  90)
  enddef;}
                \fmfleft{v1}
                \fmfright{o1,o2}
                \fmffixed{(-.2h,-.2h)}{v1,i1}
                \fmffixed{(-.2h,.2h)}{v1,i2}
                \fmffixed{(2h,0)}{v1,v2}
                \fmffixed{(1h,.4h)}{v1,v3}
                \fmffixed{(1h,1.2h)}{v1,v4}
                \fmf{tfermion,right=.2}{i1,v1,i2}
                \fmf{tfermion}{o1,v2,o2}
                \fmf{photon,tension=0,label=$\downarrow k$,label.side=left}{v4,v3}
                \fmf{plain,right=.5,label=$p\rightarrow$,label.side=right}{v1,v2}
                \fmf{plain,left=.25,label=$\stackrel{p+k}{\leftarrow}$,label.side=left}{v1,v3}
                \fmf{plain,right=.25}{v2,v3}
            \end{fmfgraph*}
        \end{tabular}
\end{fmffile}}~~ & \propto & \int d^3 p \left(2p^\m+k^{\m}\right)k^{\n}\epsilon_{\m\n\rho}\frac{1}{p^2 }= 0
\eea

As a result, only six diagrams remain to be computed. We first compute the common loop-corrected Chern-Simons gauge propagator, and record a integral formula which will be frequently used in the following calculations.
\ie
&\parbox{27mm}{\begin{fmffile}{lcpa}
        \begin{tabular}{c}
            \begin{fmfgraph*}(27,15)
                \fmfstraight
                \fmfset{arrow_len}{.3cm}\fmfset{arrow_ang}{12}
                \fmfleft{i}
                \fmfright{o}
                \fmf{photon}{i,v1}
                \fmf{photon}{o,v1}
                \fmfblob{.2w}{v1}
            \end{fmfgraph*}
        \end{tabular}
\end{fmffile}}  =  \frac{1}{32k}\left(\frac{k^{\mu}k^{\nu}}{k^{2}}-g^{\mu\nu}\right) \equiv -\frac{1}{32k}P^{\mu\nu},\label{lcplb}\\
&\int\frac{d^{3}q}{(2\pi)^{3}}\frac{(p+k)^{\mu}q^{\n}}{q^{2}(q+k)^{4}}  = -\frac{1}{32k}P^{\mu\nu}.
\fe
Here $P^{\m\n}$ is a projection operator that satisfies
\ie
&g_{\mu\nu}P^{\alpha\mu}P^{\nu\beta}  =  P^{\alpha\beta}, ~~~~~P^{\alpha\beta}\gamma_{\alpha}\gamma_{\beta} =  2.\label{lcple}
\fe
Using (\ref{lcplb})-(\ref{lcple}), we have
\be
\begin{aligned}
A_5 = \parbox{40mm}{\begin{fmffile}{A5}
        \begin{tabular}{c}
            \begin{fmfgraph*}(45,25)
 \fmfcmd{
   vardef bar (expr p, len, ang) =
    ((-len/2,0)--(len/2,0))
       rotated (ang + angle
         direction length(p)/2 of p)
       shifted point length(p)/2 of p
   enddef;
   style_def tfermion expr p =
    draw p;
   ccutdraw bar (p, 2mm,  90)
  enddef;}
                \fmfleft{i1,i2}
                \fmfright{o1,o2}
                \fmf{plain}{i2,v1}
                \fmf{tfermion}{i1,v1}
                \fmf{tfermion}{o1,v2}
                \fmf{plain}{o2,v2}
                \fmffixed{(.6h,0)}{v1,v3}
                \fmffixed{(.6h,0)}{v3,v2}
               \fmffixed{(-.3h,.25h)}{v3,v4}
                \fmffixed{(.3h,.25h)}{v3,v7}
                \fmffixed{(0,.5h)}{v3,v5}
                \fmf{plain,right=.3,tension=.3}{v3,v4,v1}
                \fmf{tfermion,right=.7,tension=.3}{v3,v2}
                \fmf{tfermion,left=.7,tension=.3}{v3,v1}
                \fmf{plain,left=.3,tension=.3}{v3,v7,v2}
                \fmf{photon,left=.3,tension=0}{v4,v5}
                \fmfblob{.1w}{v5}
                \fmf{photon,left=.3,tension=0}{v5,v7}
             \end{fmfgraph*}
        \end{tabular}
        \end{fmffile}} & = \left(-\frac{2}{32}P_{\alpha\nu}\right)\left(-\frac{2}{32}P_{\beta\mu}\right)\left(-\frac{1}{32}P^{\mu\nu}\right)\gamma^{\alpha}\gamma^{\beta} = -\frac{1}{4096},
\end{aligned}
\ee
\be
\begin{aligned}
\parbox{40mm}{\begin{fmffile}{A6}
        \begin{tabular}{c}
            \begin{fmfgraph*}(45,25)
 \fmfcmd{
   vardef bar (expr p, len, ang) =
    ((-len/2,0)--(len/2,0))
       rotated (ang + angle
         direction length(p)/2 of p)
       shifted point length(p)/2 of p
   enddef;
   style_def tfermion expr p =
    draw p;
   ccutdraw bar (p, 2mm,  90)
  enddef;}
                \fmfleft{i1,i2}
                \fmfright{o1,o2}
                \fmf{plain}{i2,v1}
                \fmf{tfermion}{i1,v1}
                \fmf{tfermion}{o1,v2}
                \fmf{plain}{o2,v2}
                \fmffixed{(.6h,0)}{v1,v3}
                \fmffixed{(.6h,0)}{v3,v2}
               \fmffixed{(-.3h,.25h)}{v3,v4}
                \fmffixed{(.3h,.25h)}{v3,v7}
                \fmffixed{(0,.5h)}{v3,v5}
                \fmf{tfermion,right=.3,tension=.3}{v3,v4,v1}
                \fmf{plain,right=.7,tension=.3}{v3,v2}
                \fmf{plain,left=.7,tension=.3}{v3,v1}
                \fmf{tfermion,left=.3,tension=.3}{v3,v7,v2}
                \fmf{photon,left=.3,tension=0}{v4,v5}
                \fmfblob{.1w}{v5}
                \fmf{photon,left=.3,tension=0}{v5,v7}
             \end{fmfgraph*}
        \end{tabular}
        \end{fmffile}} & = \left(-\frac{1}{32}P_{\alpha\beta}\right)\left(-\frac{1}{32}P_{\rho\sigma}\right)\left(-\frac{1}{32}P_{\mu\nu}\right)\gamma^{\alpha}\gamma^{\mu}\gamma^{\beta}\gamma^{\rho}\gamma^{\nu}\gamma^{\sigma} = 0,
\end{aligned}
\ee
\be
\begin{aligned}
\parbox{40mm}{\begin{fmffile}{A7}
        \begin{tabular}{c}
            \begin{fmfgraph*}(45,25)
 \fmfcmd{
   vardef bar (expr p, len, ang) =
    ((-len/2,0)--(len/2,0))
       rotated (ang + angle
         direction length(p)/2 of p)
       shifted point length(p)/2 of p
   enddef;
   style_def tfermion expr p =
    draw p;
   ccutdraw bar (p, 2mm,  90)
  enddef;}
                \fmfleft{i1,i2}
                \fmfright{o1,o2}
                \fmf{plain}{i2,v1}
                \fmf{tfermion}{i1,v1}
                \fmf{tfermion}{o1,v2}
                \fmf{plain}{o2,v2}
                \fmffixed{(.6h,0)}{v1,v3}
                \fmffixed{(.6h,0)}{v3,v2}
               \fmffixed{(-.3h,.25h)}{v3,v4}
                \fmffixed{(.3h,.25h)}{v3,v7}
                \fmffixed{(0,.5h)}{v3,v5}
                \fmf{plain,right=.3,tension=.3}{v3,v4,v1}
                \fmf{plain,right=.7,tension=.3}{v3,v2}
                \fmf{tfermion,left=.7,tension=.3}{v3,v1}
                \fmf{tfermion,left=.3,tension=.3}{v3,v7,v2}
                \fmf{photon,left=.3,tension=0}{v4,v5}
                \fmfblob{.1w}{v5}
                \fmf{photon,left=.3,tension=0}{v5,v7}
             \end{fmfgraph*}
        \end{tabular}
        \end{fmffile}} & = \left(-\frac{2}{32}P_{\alpha\mu}\right)\left(\frac{1}{32}P_{\beta\rho}\right)\left(-\frac{1}{32}P^{\mu\nu}\right)\gamma^{\alpha}\gamma^{\beta}\gamma_{\nu}\gamma^{\rho} = 0,
\end{aligned}
\ee
\be
\begin{aligned}
B_7 = \parbox{40mm}{\begin{fmffile}{B7}
        \begin{tabular}{c}
            \begin{fmfgraph*}(45,25)
 \fmfcmd{
   vardef bar (expr p, len, ang) =
    ((-len/2,0)--(len/2,0))
       rotated (ang + angle
         direction length(p)/2 of p)
       shifted point length(p)/2 of p
   enddef;
   style_def tfermion expr p =
    draw p;
   ccutdraw bar (p, 2mm,  90)
  enddef;}
                \fmfleft{i1,i2}
                \fmfright{o1,o2}
                \fmf{tfermion}{i2,v4}
                \fmf{tfermion}{v4,v1}
                \fmf{plain}{i1,v1}
                \fmf{tfermion}{o1,v2}
                \fmf{plain}{o2,v2}
                \fmffixed{(.6h,0)}{v1,v3}
                \fmffixed{(.6h,0)}{v3,v2}
               %\fmffixed{(-.3h,.25h)}{v3,v4}
                \fmffixed{(.3h,.25h)}{v3,v7}
                \fmffixed{(-.25h,.5h)}{v3,v5}
                \fmf{plain,right=.7,tension=.3}{v3,v1}
                \fmf{tfermion,right=.7,tension=.3}{v3,v2}
                \fmf{tfermion,left=.7,tension=.3}{v3,v1}
                \fmf{plain,left=.3,tension=.3}{v3,v7,v2}
                \fmf{photon,left=.1,tension=0}{v4,v5}
                \fmfblob{.1w}{v5}
                \fmf{photon,left=.1,tension=0}{v5,v7}
             \end{fmfgraph*}
        \end{tabular}
        \end{fmffile}} & = \left(-\frac{2}{32}P_{\alpha\mu}\right)\left(-\frac{1}{32}P^{\mu}_{\nu}\right)\gamma^{\n}\gamma^{\alpha}{A_{fs}}(2,2) = \frac{1}{4096},
\end{aligned}
\ee
\be
\begin{aligned}
\parbox{40mm}{\begin{fmffile}{B8}
        \begin{tabular}{c}
            \begin{fmfgraph*}(45,25)
 \fmfcmd{
   vardef bar (expr p, len, ang) =
    ((-len/2,0)--(len/2,0))
       rotated (ang + angle
         direction length(p)/2 of p)
       shifted point length(p)/2 of p
   enddef;
   style_def tfermion expr p =
    draw p;
   ccutdraw bar (p, 2mm,  90)
  enddef;}
                \fmfleft{i1,i2}
                \fmfright{o1,o2}
                \fmf{tfermion}{i2,v4}
                \fmf{tfermion}{v4,v1}
                \fmf{plain}{i1,v1}
                \fmf{tfermion}{o1,v2}
                \fmf{plain}{o2,v2}
                \fmffixed{(.6h,0)}{v1,v3}
                \fmffixed{(.6h,0)}{v3,v2}
               %\fmffixed{(-.3h,.25h)}{v3,v4}
                \fmffixed{(.3h,.25h)}{v3,v7}
                \fmffixed{(-.25h,.5h)}{v3,v5}
                \fmf{plain,right=.7,tension=.3}{v3,v1}
                \fmf{plain,right=.7,tension=.3}{v3,v2}
                \fmf{tfermion,left=.7,tension=.3}{v3,v1}
                \fmf{tfermion,left=.3,tension=.3}{v3,v7,v2}
                \fmf{photon,left=.1,tension=0}{v4,v5}
                \fmfblob{.1w}{v5}
                \fmf{photon,left=.1,tension=0}{v5,v7}
             \end{fmfgraph*}
        \end{tabular}
        \end{fmffile}} & = \left(-\frac{1}{32}P_{\alpha\beta}\right)\left(-\frac{1}{32}P_{\mu\nu}\right)\gamma^{\m}\gamma^{\beta}\gamma^{\n}\gamma^{\alpha}{A_{fs}}(2,2) = 0,
\end{aligned}
\ee
\be
\begin{aligned}
C_5 = \parbox{40mm}{\begin{fmffile}{C5}
        \begin{tabular}{c}
            \begin{fmfgraph*}(45,25)
 \fmfcmd{
   vardef bar (expr p, len, ang) =
    ((-len/2,0)--(len/2,0))
       rotated (ang + angle
         direction length(p)/2 of p)
       shifted point length(p)/2 of p
   enddef;
   style_def tfermion expr p =
    draw p;
   ccutdraw bar (p, 2mm,  90)
  enddef;}
                \fmfleft{i1,i2}
                \fmfright{o1,o2}
                \fmf{tfermion}{i2,v4}
                \fmf{tfermion}{v4,v1}
                \fmf{plain}{i1,v1}
                \fmf{plain}{o1,v2}
                \fmf{tfermion}{o2,v7,v2}
                \fmffixed{(.6h,0)}{v1,v3}
                \fmffixed{(.6h,0)}{v3,v2}
               %\fmffixed{(-.3h,.25h)}{v3,v4}
               % \fmffixed{(.3h,.25h)}{v3,v7}
                \fmffixed{(0,.5h)}{v3,v5}
                \fmf{plain,right=.7,tension=.3}{v3,v1}
                \fmf{tfermion,right=.7,tension=.3}{v3,v2}
                \fmf{tfermion,left=.7,tension=.3}{v3,v1}
                \fmf{plain,left=.7,tension=.3}{v3,v2}
                \fmf{photon,left=.1,tension=0}{v4,v5}
                \fmfblob{.1w}{v5}
                \fmf{photon,left=.1,tension=0}{v5,v7}
             \end{fmfgraph*}
        \end{tabular}
        \end{fmffile}} & = \left(-\frac{1}{32}P_{\mu\nu}\right)\gamma^{\m}\gamma^{\n}{A_{fs}}(2,2)^2 = -\frac{1}{4096}.
\end{aligned}
\ee
Once again we have omitted the common factor (\ref{common_loop}).

Finally, we need to put in the factors associated with vertices and gauge/flavor indices. The relevant vertices are listed as follows.
%\bigskip

\bea
\parbox{15mm}{\begin{fmffile}{Dvertex1}
        \begin{tabular}{c}
            \begin{fmfgraph*}(15,15)
 \fmfcmd{
   vardef bar (expr p, len, ang) =
    ((-len/2,0)--(len/2,0))
       rotated (ang + angle
         direction length(p)/2 of p)
       shifted point length(p)/2 of p
   enddef;
   style_def tfermion expr p =
    draw p;
   ccutdraw bar (p, 2mm,  90)
  enddef;}
                \fmfleft{i1,i2}
                \fmfright{o1,o2}
                \fmflabel{$\phi^i$}{i1}
                \fmflabel{$\bar\phi^i$}{i2}
                \fmflabel{$\psi^j$}{o1}
                \fmflabel{$\bar\psi^j$}{o2}
                \fmf{plain}{i1,d,i2}
                \fmf{tfermion}{o1,d,o2}
                \fmfdot{d}
            \end{fmfgraph*}
        \end{tabular}
        \end{fmffile}}
& = & \frac{2\pi}{k}~~~~~~
\parbox{15mm}{\begin{fmffile}{Dvertex2}
        \begin{tabular}{c}
            \begin{fmfgraph*}(15,15)
 \fmfcmd{
   vardef bar (expr p, len, ang) =
    ((-len/2,0)--(len/2,0))
       rotated (ang + angle
         direction length(p)/2 of p)
       shifted point length(p)/2 of p
   enddef;
   style_def tfermion expr p =
    draw p;
   ccutdraw bar (p, 2mm,  90)
  enddef;}
                \fmfleft{i1,i2}
                \fmfright{o1,o2}
                \fmflabel{$\phi^i$}{i1}
                \fmflabel{$\bar\phi^i$}{i2}
                \fmflabel{$\bar\psi^j$}{o1}
                \fmflabel{$\psi^j$}{o2}
                \fmf{plain}{i1,d,i2}
                \fmf{tfermion}{o2,d,o1}
                \fmfdot{d}
            \end{fmfgraph*}
        \end{tabular}
        \end{fmffile}}
 =  -\frac{2\pi}{k}~~~~~~
\parbox{15mm}{\begin{fmffile}{Dvertex3}
        \begin{tabular}{c}
            \begin{fmfgraph*}(15,15)
 \fmfcmd{
   vardef bar (expr p, len, ang) =
    ((-len/2,0)--(len/2,0))
       rotated (ang + angle
         direction length(p)/2 of p)
       shifted point length(p)/2 of p
   enddef;
   style_def tfermion expr p =
    draw p;
   ccutdraw bar (p, 2mm,  90)
  enddef;}
                \fmfleft{i1,i2}
                \fmfright{o1,o2}
                \fmflabel{$\phi^i$}{i1}
                \fmflabel{$\psi^j$}{i2}
                \fmflabel{$\bar\psi^i$}{o1}
                \fmflabel{$\bar\phi^j$}{o2}
                \fmf{plain}{i1,d,o2}
                \fmf{tfermion}{i2,d,o1}
                \fmfdot{d}
            \end{fmfgraph*}
        \end{tabular}
        \end{fmffile}}
 =  \frac{4\pi}{k}~~~~~
\parbox{15mm}{\begin{fmffile}{Dvertex4}
        \begin{tabular}{c}
            \begin{fmfgraph*}(15,15)
 \fmfcmd{
   vardef bar (expr p, len, ang) =
    ((-len/2,0)--(len/2,0))
       rotated (ang + angle
         direction length(p)/2 of p)
       shifted point length(p)/2 of p
   enddef;
   style_def tfermion expr p =
    draw p;
   ccutdraw bar (p, 2mm,  90)
  enddef;}
                \fmfleft{i1,i2}
                \fmfright{o1,o2}
                \fmflabel{$\phi^i$}{i1}
                \fmflabel{$\bar\psi^i$}{i2}
                \fmflabel{$\bar\phi^j$}{o1}
                \fmflabel{$\psi^j$}{o2}
                \fmf{plain}{i1,d,o1}
                \fmf{tfermion}{o2,d,i2}
                \fmfdot{d}
            \end{fmfgraph*}
        \end{tabular}
        \end{fmffile}}
 =  -\frac{4\pi}{k}
\label{Dvertex}
\eea

%\bigskip
\bigskip

\bea
\parbox{20mm}{\begin{fmffile}{Dvertex5}
        \begin{tabular}{c}
            \begin{fmfgraph*}(15,15)
                \fmfleft{i1,i2}
                \fmfright{o}
                \fmflabel{$\phi^i$}{i1}
                \fmflabel{$\bar\phi^i$}{i2}
                \fmflabel{$A_\m$}{o}
                \fmf{plain}{i1,d,i2}
                \fmf{photon}{o,d}
                \fmfdot{d}
            \end{fmfgraph*}
        \end{tabular}
       \end{fmffile}}~  =  \sqrt{\frac{4\pi}{k}}\left(2p_\m+k_\m\right)~~~~~~~
\parbox{20mm}{\begin{fmffile}{Dvertex6}
        \begin{tabular}{c}
            \begin{fmfgraph*}(15,15)
 \fmfcmd{
   vardef bar (expr p, len, ang) =
    ((-len/2,0)--(len/2,0))
       rotated (ang + angle
         direction length(p)/2 of p)
       shifted point length(p)/2 of p
   enddef;
   style_def tfermion expr p =
    draw p;
   ccutdraw bar (p, 2mm,  90)
  enddef;}
                \fmfleft{i1,i2}
                \fmfright{o}
                \fmflabel{$\psi^i$}{i1}
                \fmflabel{$\bar\psi^i$}{i2}
                \fmflabel{$A_{\m}$}{o}
                \fmf{tfermion}{i1,d,i2}
                \fmf{photon}{o,d}
                \fmfdot{d}
            \end{fmfgraph*}
        \end{tabular}
       \end{fmffile}}~  =  \sqrt{\frac{4\pi}{k}}\gamma_\m
\eea

%\bigskip
\bigskip

\bea
\parbox{15mm}{\begin{fmffile}{Fvertex1}
        \begin{tabular}{c}
            \begin{fmfgraph*}(15,15)
 \fmfcmd{
   vardef bar (expr p, len, ang) =
    ((-len/2,0)--(len/2,0))
       rotated (ang + angle
         direction length(p)/2 of p)
       shifted point length(p)/2 of p
   enddef;
   style_def tfermion expr p =
    draw p;
   ccutdraw bar (p, 2mm,  90)
  enddef;}
                \fmfleft{i1,i2}
                \fmfright{o1,o2}
                \fmflabel{$\phi^i$}{i1}
                \fmflabel{$\psi^l$}{i2}
                \fmflabel{$\phi^j$}{o1}
                \fmflabel{$\psi^k$}{o2}
                \fmf{plain}{i1,d}
                \fmf{tfermion}{i2,d}
                \fmf{plain}{o1,d}
                \fmf{tfermion}{o2,d}
                \fmfdot{d}
            \end{fmfgraph*}
        \end{tabular}
        \end{fmffile}}
 =  -\alpha_{ijkl}~~~~~
\parbox{15mm}{\begin{fmffile}{Fvertex2}
        \begin{tabular}{c}
            \begin{fmfgraph*}(15,15)
 \fmfcmd{
   vardef bar (expr p, len, ang) =
    ((-len/2,0)--(len/2,0))
       rotated (ang + angle
         direction length(p)/2 of p)
       shifted point length(p)/2 of p
   enddef;
   style_def tfermion expr p =
    draw p;
   ccutdraw bar (p, 2mm,  90)
  enddef;}
                \fmfleft{i1,i2}
                \fmfright{o1,o2}
                \fmflabel{$\phi^i$}{i1}
                \fmflabel{$\psi^l$}{i2}
                \fmflabel{$\psi^j$}{o1}
                \fmflabel{$\phi^k$}{o2}
                \fmf{plain}{i1,d}
                \fmf{tfermion}{i2,d}
                \fmf{tfermion}{o1,d}
                \fmf{plain}{o2,d}
                \fmfdot{d}
            \end{fmfgraph*}
        \end{tabular}
        \end{fmffile}}
 =  -\alpha_{ijkl}~~~~~
\parbox{15mm}{\begin{fmffile}{Fvertex3}
        \begin{tabular}{c}
            \begin{fmfgraph*}(15,15)
           \fmfcmd{
   vardef fbar (expr p, len, ang) =
    ((-len/2,0)--(len/2,0))
       rotated (ang + angle
         direction length(p)/2 of p)
       shifted point length(p)*45/100 of p
   enddef;
   vardef bbar (expr p, len, ang) =
    ((-len/2,0)--(len/2,0))
       rotated (ang + angle
         direction length(p)/2 of p)
       shifted point length(p)*55/100 of p
   enddef;
   style_def ttfermion expr p =
    draw p;
   ccutdraw fbar (p, 2mm,  90);
   ccutdraw bbar (p, 2mm,  90)
  enddef;}
                \fmfleft{i1,i2}
                \fmfright{o1,o2}
                \fmflabel{$\phi^i$}{i1}
                \fmflabel{$\phi^l$}{i2}
                \fmflabel{$\phi^j$}{o1}
                \fmflabel{$F^k$}{o2}
                \fmf{plain}{i1,d}
                \fmf{plain}{i2,d}
                \fmf{plain}{o1,d}
                \fmf{ttfermion}{o2,d}
                \fmfdot{d}
            \end{fmfgraph*}
        \end{tabular}
        \end{fmffile}}
 =  \alpha_{ijkl}
\eea

\bigskip

\noindent
where the ``D-term" vertices in the first row are obtained after integrating out the auxiliary fields. Each diagram also comes with a common factor $(\frac{2\pi}{k})^2 N^4 M\left(-\alpha_{ijmn}\right)\left(-\bar\alpha^{mnpq}\right)\left(-\alpha_{pqkl}\right)$, which has been taken into account in (\ref{potential}).

The above results allow us to compute the coefficients of the last four terms in (\ref{potential}). Here we evaluate explicitly the beta function coefficient for the Yukawa coupling ${\rm Tr}\left(\psi^i\psi^j\phi^k\phi^l\right)$; the other coefficients are computed in a similar manner.
\bea
2\parbox{27mm}{\begin{fmffile}{A2a}
        \begin{tabular}{c}
            \begin{fmfgraph*}(27,15)
       \fmfcmd{
   vardef bar (expr p, len, ang) =
    ((-len/2,0)--(len/2,0))
       rotated (ang + angle
         direction length(p)/2 of p)
       shifted point length(p)/2 of p
   enddef;
   style_def tfermion expr p =
    draw p;
   ccutdraw bar (p, 2mm,  90)
  enddef;}
         \fmfcmd{
   vardef bar (expr p, len, ang) =
    ((-len/2,0)--(len/2,0))
       rotated (ang + angle
         direction length(p)/2 of p)
       shifted point length(p)/2 of p
   enddef;
   style_def dashfermion expr p =
    draw_dashes p;
   ccutdraw bar (p, 2mm,  90)
  enddef;}
                \fmfleft{i1,i2}
                \fmfright{o1,o2}
                \fmf{tfermion}{i2,v1}
                \fmf{tfermion}{i1,v1}
                \fmf{plain}{v2,o2}
                \fmf{plain}{v2,o1}
                \fmffixed{(.6h,0)}{v1,v3}
                \fmffixed{(.6h,0)}{v3,v2}
                \fmffixed{(-.3h,.25h)}{v3,v4}
                \fmffixed{(.3h,.25h)}{v3,v5}
                \fmf{plain,right=.7,tension=.3}{v1,v3}
                \fmf{tfermion,right=.7,tension=.3}{v3,v2}
                \fmf{plain,left=.3,tension=.3}{v1,v4}
                \fmf{tfermion,left=.3,tension=.3}{v4,v3}
                \fmf{plain,left=.3,tension=.3}{v3,v5}
                \fmf{tfermion,left=.3,tension=.3}{v5,v2}
                \fmf{dashfermion,left=.7,tension=0}{v4,v5}
             \end{fmfgraph*}
        \end{tabular}
        \end{fmffile}} & = & 2A_2\\
2\left[\parbox{27mm}{\begin{fmffile}{B3a}
        \begin{tabular}{c}
            \begin{fmfgraph*}(27,15)
       \fmfcmd{
   vardef bar (expr p, len, ang) =
    ((-len/2,0)--(len/2,0))
       rotated (ang + angle
         direction length(p)/2 of p)
       shifted point length(p)/2 of p
   enddef;
   style_def tfermion expr p =
    draw p;
   ccutdraw bar (p, 2mm,  90)
  enddef;}
           \fmfcmd{
   vardef bar (expr p, len, ang) =
    ((-len/2,0)--(len/2,0))
       rotated (ang + angle
         direction length(p)/2 of p)
       shifted point length(p)/2 of p
   enddef;
   style_def dashfermion expr p =
    draw_dashes p;
   ccutdraw bar (p, 2mm,  90)
  enddef;}
                \fmfleft{i1,i2}
                \fmfright{o1,o2}
                \fmf{tfermion}{i2,v4}
                \fmf{plain}{v4,v1}
                \fmf{tfermion}{i1,v1}
                \fmf{plain}{v2,o2}
                \fmf{plain}{v2,o1}
                \fmffixed{(.6h,0)}{v1,v3}
                \fmffixed{(.6h,0)}{v3,v2}
               %\fmffixed{(-.3h,.25h)}{v3,v4}
                \fmffixed{(.3h,.25h)}{v3,v5}
                \fmf{tfermion,right=.7,tension=.3}{v1,v3}
                \fmf{tfermion,right=.7,tension=.3}{v3,v2}
                \fmf{plain,left=.7,tension=.3}{v1,v3}
                \fmf{plain,left=.3,tension=.3}{v3,v5}
                \fmf{tfermion,left=.3,tension=.3}{v5,v2}
                \fmf{dashfermion,left=.35,tension=0}{v4,v5}
             \end{fmfgraph*}
        \end{tabular}
        \end{fmffile}}+
\parbox{27mm}{\begin{fmffile}{B3b}
        \begin{tabular}{c}
            \begin{fmfgraph*}(27,15)
       \fmfcmd{
   vardef bar (expr p, len, ang) =
    ((-len/2,0)--(len/2,0))
       rotated (ang + angle
         direction length(p)/2 of p)
       shifted point length(p)/2 of p
   enddef;
   style_def tfermion expr p =
    draw p;
   ccutdraw bar (p, 2mm,  90)
  enddef;}
           \fmfcmd{
   vardef bar (expr p, len, ang) =
    ((-len/2,0)--(len/2,0))
       rotated (ang + angle
         direction length(p)/2 of p)
       shifted point length(p)/2 of p
   enddef;
   style_def dashfermion expr p =
    draw_dashes p;
   ccutdraw bar (p, 2mm,  90)
  enddef;}
                \fmfleft{i1,i2}
                \fmfright{o1,o2}
                \fmf{tfermion}{i2,v4}
                \fmf{plain}{v4,v1}
                \fmf{tfermion}{i1,v1}
                \fmf{plain}{v2,o2}
                \fmf{plain}{v2,o1}
                \fmffixed{(.6h,0)}{v1,v3}
                \fmffixed{(.6h,0)}{v3,v2}
               %\fmffixed{(-.3h,.25h)}{v3,v4}
                \fmffixed{(.3h,.25h)}{v3,v5}
                \fmf{plain,right=.7,tension=.3}{v1,v3}
                \fmf{tfermion,right=.7,tension=.3}{v3,v2}
                \fmf{tfermion,left=.7,tension=.3}{v1,v3}
                \fmf{plain,left=.3,tension=.3}{v3,v5}
                \fmf{tfermion,left=.3,tension=.3}{v5,v2}
                \fmf{dashfermion,left=.35,tension=0}{v4,v5}
             \end{fmfgraph*}
        \end{tabular}
        \end{fmffile}}
\right] & = & 4B_3\\
2\left[\parbox{27mm}{\begin{fmffile}{B4a}
        \begin{tabular}{c}
            \begin{fmfgraph*}(27,15)
       \fmfcmd{
   vardef bar (expr p, len, ang) =
    ((-len/2,0)--(len/2,0))
       rotated (ang + angle
         direction length(p)/2 of p)
       shifted point length(p)/2 of p
   enddef;
   style_def tfermion expr p =
    draw p;
   ccutdraw bar (p, 2mm,  90)
  enddef;}
           \fmfcmd{
   vardef bar (expr p, len, ang) =
    ((-len/2,0)--(len/2,0))
       rotated (ang + angle
         direction length(p)/2 of p)
       shifted point length(p)/2 of p
   enddef;
   style_def dashfermion expr p =
    draw_dashes p;
   ccutdraw bar (p, 2mm,  90)
  enddef;}
                \fmfleft{o1,o2}
                \fmfright{i1,i2}
                \fmf{plain}{i2,v4}
                \fmf{tfermion}{v4,v1}
                \fmf{plain}{i1,v1}
                \fmf{tfermion}{v2,o2}
                \fmf{tfermion}{v2,o1}
                \fmffixed{(-.6h,0)}{v1,v3}
                \fmffixed{(-.6h,0)}{v3,v2}
               %\fmffixed{(-.3h,.25h)}{v3,v4}
                \fmffixed{(-.3h,.25h)}{v3,v5}
                \fmf{plain,left=.7,tension=.3}{v1,v3}
                \fmf{plain,left=.7,tension=.3}{v3,v2}
                \fmf{tfermion,right=.7,tension=.3}{v1,v3}
                \fmf{tfermion,right=.3,tension=.3}{v3,v5}
                \fmf{plain,right=.3,tension=.3}{v5,v2}
                \fmf{dashfermion,right=.35,tension=0}{v4,v5}
             \end{fmfgraph*}
        \end{tabular}
        \end{fmffile}}+\parbox{27mm}{\begin{fmffile}{B4b}
        \begin{tabular}{c}
            \begin{fmfgraph*}(27,15)
       \fmfcmd{
   vardef bar (expr p, len, ang) =
    ((-len/2,0)--(len/2,0))
       rotated (ang + angle
         direction length(p)/2 of p)
       shifted point length(p)/2 of p
   enddef;
   style_def tfermion expr p =
    draw p;
   ccutdraw bar (p, 2mm,  90)
  enddef;}
           \fmfcmd{
   vardef bar (expr p, len, ang) =
    ((-len/2,0)--(len/2,0))
       rotated (ang + angle
         direction length(p)/2 of p)
       shifted point length(p)/2 of p
   enddef;
   style_def dashfermion expr p =
    draw_dashes p;
   ccutdraw bar (p, 2mm,  90)
  enddef;}
                \fmfleft{o1,o2}
                \fmfright{i1,i2}
                \fmf{plain}{i2,v4}
                \fmf{tfermion}{v4,v1}
                \fmf{plain}{i1,v1}
                \fmf{tfermion}{v2,o2}
                \fmf{tfermion}{v2,o1}
                \fmffixed{(-.6h,0)}{v1,v3}
                \fmffixed{(-.6h,0)}{v3,v2}
               %\fmffixed{(-.3h,.25h)}{v3,v4}
                \fmffixed{(-.3h,.25h)}{v3,v5}
                \fmf{tfermion,left=.7,tension=.3}{v1,v3}
                \fmf{plain,left=.7,tension=.3}{v3,v2}
                \fmf{plain,right=.7,tension=.3}{v1,v3}
                \fmf{tfermion,right=.3,tension=.3}{v3,v5}
                \fmf{plain,right=.3,tension=.3}{v5,v2}
                \fmf{dashfermion,right=.35,tension=0}{v4,v5}
             \end{fmfgraph*}
        \end{tabular}
        \end{fmffile}}\right] & =& 4B_4\\
2\left[
\parbox{27mm}{\begin{fmffile}{C4a}
        \begin{tabular}{c}
            \begin{fmfgraph*}(27,15)
       \fmfcmd{
   vardef bar (expr p, len, ang) =
    ((-len/2,0)--(len/2,0))
       rotated (ang + angle
         direction length(p)/2 of p)
       shifted point length(p)/2 of p
   enddef;
   style_def tfermion expr p =
    draw p;
   ccutdraw bar (p, 2mm,  90)
  enddef;}
           \fmfcmd{
   vardef bar (expr p, len, ang) =
    ((-len/2,0)--(len/2,0))
       rotated (ang + angle
         direction length(p)/2 of p)
       shifted point length(p)/2 of p
   enddef;
   style_def dashfermion expr p =
    draw_dashes p;
   ccutdraw bar (p, 2mm,  90)
  enddef;}
                \fmfleft{i1,i2}
                \fmfright{o1,o2}
                \fmf{tfermion}{i2,v4}
                \fmf{plain}{v4,v1}
                \fmf{tfermion}{i1,v1}
                \fmf{tfermion}{v2,v5}
                \fmf{plain}{v5,o2}
                \fmf{plain}{v2,o1}
                \fmffixed{(.6h,0)}{v1,v3}
                \fmffixed{(.6h,0)}{v3,v2}
               %\fmffixed{(-.3h,.25h)}{v3,v4}
                %\fmffixed{(.3h,.25h)}{v3,v5}
                \fmf{tfermion,right=.7,tension=.3}{v1,v3}
                \fmf{tfermion,right=.7,tension=.3}{v3,v2}
                \fmf{plain,left=.7,tension=.3}{v1,v3}
                \fmf{plain,left=.7,tension=.3}{v3,v2}
                \fmf{dashfermion,left=.2,tension=0}{v4,v5}
             \end{fmfgraph*}
        \end{tabular}
        \end{fmffile}}+
\parbox{27mm}{\begin{fmffile}{C4b}
        \begin{tabular}{c}
            \begin{fmfgraph*}(27,15)
       \fmfcmd{
   vardef bar (expr p, len, ang) =
    ((-len/2,0)--(len/2,0))
       rotated (ang + angle
         direction length(p)/2 of p)
       shifted point length(p)/2 of p
   enddef;
   style_def tfermion expr p =
    draw p;
   ccutdraw bar (p, 2mm,  90)
  enddef;}
           \fmfcmd{
   vardef bar (expr p, len, ang) =
    ((-len/2,0)--(len/2,0))
       rotated (ang + angle
         direction length(p)/2 of p)
       shifted point length(p)/2 of p
   enddef;
   style_def dashfermion expr p =
    draw_dashes p;
   ccutdraw bar (p, 2mm,  90)
  enddef;}
                \fmfleft{i1,i2}
                \fmfright{o1,o2}
                \fmf{tfermion}{i2,v4}
                \fmf{plain}{v4,v1}
                \fmf{tfermion}{i1,v1}
                \fmf{tfermion}{v2,v5}
                \fmf{plain}{v5,o2}
                \fmf{plain}{v2,o1}
                \fmffixed{(.6h,0)}{v1,v3}
                \fmffixed{(.6h,0)}{v3,v2}
               %\fmffixed{(-.3h,.25h)}{v3,v4}
                %\fmffixed{(.3h,.25h)}{v3,v5}
                \fmf{plain,right=.7,tension=.3}{v1,v3}
                \fmf{tfermion,right=.7,tension=.3}{v3,v2}
                \fmf{tfermion,left=.7,tension=.3}{v1,v3}
                \fmf{plain,left=.7,tension=.3}{v3,v2}
                \fmf{dashfermion,left=.2,tension=0}{v4,v5}
             \end{fmfgraph*}
        \end{tabular}
        \end{fmffile}}+2~{\rm more}\right] & = & 8C_4
\eea
The diagrams of type $A,B,C$ correspond to the contributions of the three different supergraphs in
(\ref{ABC}). Taking into account the signs and factors of $2$ coming from the vertices, we end up with
\bea
2c_1 & = & a+b+c\\
a & = & 2A_2\times 4\times \left(\frac{1}{2\pi^2}\right)=\frac{1}{256\pi^2}\\
b & = & \left(4B_3+4B_4\right)\times 4\times (-1)\left(\frac{1}{2\pi^2}\right)=-\frac{1}{128\pi^2}\\
c & = & 8C_4\times 4\times \left(\frac{1}{2\pi^2}\right)= \frac{1}{256\pi^2}
\eea
where 
the factors of $4$ come from the relative factor of 2 between the first two vertices and the last two vertices in (\ref{Dvertex}), the minus signs of $B_3$ and $B_4$ are given by the relative minus sign of the first two vertices in (\ref{Dvertex}), and the $1/2\pi^2$ comes from (\ref{common_loop}).

\section{Graphical rules}

Now we describe some graphical rules that were used to simplify bubble diagrams in the previous section. A generalized scalar line and a generalized fermion line with a label are defined as
\bea\nonumber
\parbox{25mm}{\begin{fmffile}{gs}
        \begin{tabular}{c}
            \begin{fmfgraph*}(25,20)
                \fmfleft{i1}
                \fmfright{o1}
                \fmf{plain,label=$a$}{i1,o1}
             \end{fmfgraph*}
        \end{tabular}
        \end{fmffile}}  = \frac{1}{k^a},~~~~~~~~~~~~~ 
\parbox{25mm}{\begin{fmffile}{gf}
        \begin{tabular}{c}
            \begin{fmfgraph*}(25,20)
            \fmfcmd{
   vardef bar (expr p, len, ang) =
    ((-len/2,0)--(len/2,0))
       rotated (ang + angle
         direction length(p)/2 of p)
       shifted point length(p)/2 of p
   enddef;
   style_def tfermion expr p =
    draw p;
   ccutdraw bar (p, 3mm,  90)
  enddef;}
                \fmfleft{i1}
                \fmfright{o1}
                \fmf{tfermion,label=$a$}{i1,o1}
             \end{fmfgraph*}
        \end{tabular}
        \end{fmffile}}  =  \frac{i\slash \!\!\!k}{k^a}.
\eea
They may be combined according to
\ie
&\parbox{35mm}{\begin{fmffile}{rule1}
        \begin{tabular}{c}
            \begin{fmfgraph*}(35,20)
       \fmfcmd{
   vardef bar (expr p, len, ang) =
    ((-len/2,0)--(len/2,0))
       rotated (ang + angle
         direction length(p)/2 of p)
       shifted point length(p)/2 of p
   enddef;
   style_def tfermion expr p =
    draw p;
   ccutdraw bar (p, 2mm,  90)
  enddef;}
                \fmfleft{i1}
                \fmfright{o1}
                \fmf{plain,label=$a$}{i1,v1}
                \fmf{plain,label=$b$}{v1,o1}
                \fmfdot{i1,v1,o1}
             \end{fmfgraph*}
        \end{tabular}
        \end{fmffile}}  = 
\parbox{35mm}{\begin{fmffile}{rule1a}
        \begin{tabular}{c}
            \begin{fmfgraph*}(25,20)
       \fmfcmd{
   vardef bar (expr p, len, ang) =
    ((-len/2,0)--(len/2,0))
       rotated (ang + angle
         direction length(p)/2 of p)
       shifted point length(p)/2 of p
   enddef;
   style_def tfermion expr p =
    draw p;
   ccutdraw bar (p, 2mm,  90)
  enddef;}
                \fmfleft{i1}
                \fmfright{o1}
                \fmf{plain,label=$a+b$}{i1,o1}
                \fmfdot{i1,o1}
             \end{fmfgraph*}
        \end{tabular}
        \end{fmffile}} \\ &
\parbox{35mm}{\begin{fmffile}{rule2}
        \begin{tabular}{c}
            \begin{fmfgraph*}(35,20)
       \fmfcmd{
   vardef bar (expr p, len, ang) =
    ((-len/2,0)--(len/2,0))
       rotated (ang + angle
         direction length(p)/2 of p)
       shifted point length(p)/2 of p
   enddef;
   style_def tfermion expr p =
    draw p;
   ccutdraw bar (p, 2mm,  90)
  enddef;}
                \fmfleft{i1}
                \fmfright{o1}
                \fmf{plain,label=$a$}{i1,v1}
                \fmf{tfermion,label=$b$}{v1,o1}
                \fmfdot{i1,v1,o1}
             \end{fmfgraph*}
        \end{tabular}
        \end{fmffile}}  = 
\parbox{35mm}{\begin{fmffile}{rule2a}
        \begin{tabular}{c}
            \begin{fmfgraph*}(25,20)
       \fmfcmd{
   vardef bar (expr p, len, ang) =
    ((-len/2,0)--(len/2,0))
       rotated (ang + angle
         direction length(p)/2 of p)
       shifted point length(p)/2 of p
   enddef;
   style_def tfermion expr p =
    draw p;
   ccutdraw bar (p, 2mm,  90)
  enddef;}
                \fmfleft{i1}
                \fmfright{o1}
                \fmf{tfermion,label=$a+b$}{i1,o1}
                \fmfdot{i1,o1}
             \end{fmfgraph*}
        \end{tabular}
        \end{fmffile}}\\ &
\parbox{35mm}{\begin{fmffile}{rule3}
        \begin{tabular}{c}
            \begin{fmfgraph*}(35,20)
       \fmfcmd{
   vardef bar (expr p, len, ang) =
    ((-len/2,0)--(len/2,0))
       rotated (ang + angle
         direction length(p)/2 of p)
       shifted point length(p)/2 of p
   enddef;
   style_def tfermion expr p =
    draw p;
   ccutdraw bar (p, 2mm,  90)
  enddef;}
                \fmfleft{i1}
                \fmfright{o1}
                \fmf{tfermion,label=$a$}{i1,v1}
                \fmf{tfermion,label=$b$}{v1,o1}
                \fmfdot{i1,v1,o1}
             \end{fmfgraph*}
        \end{tabular}
        \end{fmffile}}  =  (-1)
\parbox{35mm}{\begin{fmffile}{rule3a}
        \begin{tabular}{c}
            \begin{fmfgraph*}(25,20)
       \fmfcmd{
   vardef bar (expr p, len, ang) =
    ((-len/2,0)--(len/2,0))
       rotated (ang + angle
         direction length(p)/2 of p)
       shifted point length(p)/2 of p
   enddef;
   style_def tfermion expr p =
    draw p;
   ccutdraw bar (p, 2mm,  90)
  enddef;}
                \fmfleft{i1}
                \fmfright{o1}
                \fmf{plain,label=$a+b-2$}{i1,o1}
                \fmfdot{i1,o1}
             \end{fmfgraph*}
        \end{tabular}
        \end{fmffile}}
\fe
or in bubbles,

\be
\begin{aligned}
\parbox{25mm}{\begin{fmffile}{rule4}
        \begin{tabular}{c}
            \begin{fmfgraph*}(25,20)
                \fmfleft{i1}
                \fmfright{o1}
                \fmf{plain,left=.5,label=$b$}{i1,o1}
                \fmf{plain,right=.5,label=$a$}{i1,o1}
                \fmfdot{i1,o1}
             \end{fmfgraph*}
        \end{tabular}
        \end{fmffile}} & = \int \frac{d^d q}{(2\pi)^d}\frac{1}{(q^2)^{a/2}((q-k)^2)^{b/2}} \\
& = \frac{1}{B\left(\frac{a}{2},\frac{b}{2}\right)}\int^1_0dx\int\frac{d^dq}{(2\pi)^d}\frac{x^{\frac{b}{2}}(1-x)^{\frac{a}{2}-1}}{\left[q^2+k^2x(1-x)\right]^{\frac{a+b}{2}}} \\
& \equiv {A_{ss}}(a,b)
\parbox{25mm}{\begin{fmffile}{rule4a}
        \begin{tabular}{c}
            \begin{fmfgraph*}(25,20)
                \fmfleft{i1}
                \fmfright{o1}
                \fmf{plain,label=$a+b-d$}{i1,o1}
             \end{fmfgraph*}
        \end{tabular}
        \end{fmffile}}
\end{aligned}
\ee
\be
\begin{aligned}
\parbox{25mm}{\begin{fmffile}{rule5}
        \begin{tabular}{c}
            \begin{fmfgraph*}(25,20)
       \fmfcmd{
   vardef bar (expr p, len, ang) =
    ((-len/2,0)--(len/2,0))
       rotated (ang + angle
         direction length(p)/2 of p)
       shifted point length(p)/2 of p
   enddef;
   style_def tfermion expr p =
    draw p;
   ccutdraw bar (p, 3mm,  90)
  enddef;}
                \fmfleft{i1}
                \fmfright{o1}
                \fmf{plain,left=.5,label=$b$}{i1,o1}
                \fmf{tfermion,right=.5,label=$a$}{i1,o1}
                \fmfdot{i1,o1}
             \end{fmfgraph*}
        \end{tabular}
        \end{fmffile}} & = i\int\frac{d^d q}{(2\pi)^d}\frac{\slash \!\!\!q}{(q^2)^{a/2}((q-k)^2)^{b/2}} \\
& = \frac{i}{B\left(\frac{a}{2},\frac{b}{2}\right)}\int^1_0 dx\int\frac{d^d q}{(2\pi)^d}\frac{x^{\frac{b}{2}-1}(1-x)^{\frac{a}{2}-1}\slash \!\!\!k x}{\left[q^2+k^2x(1-x)\right]^{\frac{a+b}{2}}} \\
& \equiv {A_{fs}}(a,b)
\parbox{25mm}{\begin{fmffile}{rule5a}
        \begin{tabular}{c}
            \begin{fmfgraph*}(25,20)
            \fmfcmd{
   vardef bar (expr p, len, ang) =
    ((-len/2,0)--(len/2,0))
       rotated (ang + angle
         direction length(p)/2 of p)
       shifted point length(p)/2 of p
   enddef;
   style_def tfermion expr p =
    draw p;
   ccutdraw bar (p, 3mm,  90)
  enddef;}
                \fmfleft{i1}
                \fmfright{o1}
                \fmf{tfermion,label=$a+b-d$}{i1,o1}
             \end{fmfgraph*}
        \end{tabular}
        \end{fmffile}}
\end{aligned}
\ee
\be
\begin{aligned}
\parbox{25mm}{\begin{fmffile}{rule6}
        \begin{tabular}{c}
            \begin{fmfgraph*}(25,20)
       \fmfcmd{
   vardef bar (expr p, len, ang) =
    ((-len/2,0)--(len/2,0))
       rotated (ang + angle
         direction length(p)/2 of p)
       shifted point length(p)/2 of p
   enddef;
   style_def tfermion expr p =
    draw p;
   ccutdraw bar (p, 3mm,  90)
  enddef;}
                \fmfleft{i1}
                \fmfright{o1}
                \fmf{tfermion,left=.5,label=$b$}{i1,o1}
                \fmf{tfermion,right=.5,label=$a$}{i1,o1}
                \fmfdot{i1,o1}
             \end{fmfgraph*}
        \end{tabular}
        \end{fmffile}}
 & = -i^2\int\frac{d^d q}{(2\pi)^d}\frac{{\rm Tr}[\slash \!\!\!q(\slash \!\!\!q-\slash \!\!\! k)]}{(q^2)^{a/2}((q-k)^2)^{b/2}} \\
& =  \frac{i}{B\left(\frac{a}{2},\frac{b}{2}\right)}\int^1_0 dx\int\frac{d^d q}{(2\pi)^d}\frac{2x^{\frac{b}{2}-1}(1-x)^{\frac{a}{2}-1}\left[q^2-k^2 x(1-x)\right]}{\left[q^2+k^2x(1-x)\right]^{\frac{a+b}{2}}} \\
& \equiv  {A_{ff}}(a,b)
\parbox{25mm}{\begin{fmffile}{rule6a}
        \begin{tabular}{c}
            \begin{fmfgraph*}(25,20)
                \fmfleft{i1}
                \fmfright{o1}
                \fmf{plain,label=$a+b-2-d$}{i1,o1}
             \end{fmfgraph*}
        \end{tabular}
        \end{fmffile}}
\label{rule6}
\end{aligned}
\ee
where the coefficients $A_{ss}, A_{fs}, A_{ff}$ are given by
\ie
&{A_{ss}}(a,b) = \frac{(4\pi)^{-\frac{d}{2}}\Gamma\left(\frac{d-a}{2}\right)\Gamma\left(\frac{d-b}{2}\right)\Gamma\left(\frac{a+b-d}{2}\right)}{\Gamma\left(\frac{a}{2}\right)\Gamma\left(\frac{b}{2}\right)\Gamma\left(d-\frac{a+b}{2}\right)},\\
&{A_{fs}}(a,b) = \frac{(4\pi)^{-\frac{d}{2}}\Gamma\left(\frac{d-a}{2}+1\right)\Gamma\left(\frac{d-b}{2}\right)\Gamma\left(\frac{a+b-d}{2}\right)}{\Gamma\left(\frac{a}{2}\right)\Gamma\left(\frac{b}{2}\right)\Gamma\left(d-\frac{a+b}{2}+1\right)},\\
&{A_{ff}}(a,b)  =  \frac{(4\pi)^{-\frac{d}{2}}\Gamma\left(\frac{d-a}{2}+1\right)\Gamma\left(\frac{d-b}{2}+1\right)\Gamma\left(\frac{a+b-d}{2}-1\right)}{\Gamma\left(\frac{a}{2}\right)\Gamma\left(\frac{b}{2}\right)\Gamma\left(d-\frac{a+b}{2}+1\right)}.
\fe
In the last rule (\ref{rule6}), we have assumed a fermion loop, thus traced over the spinor indices and added an overall minus sign. In general, when applying the graphical rules repeatedly, we may also encounter ``untraced" structure like
\ie
\parbox{40mm}{\begin{fmffile}{ex4}
        \begin{tabular}{c}
            \begin{fmfgraph*}(30,40)
       \fmfcmd{
   vardef bar (expr p, len, ang) =
    ((-len/2,0)--(len/2,0))
       rotated (ang + angle
         direction length(p)/2 of p)
       shifted point length(p)/2 of p
   enddef;
   style_def tfermion expr p =
    draw p;
   ccutdraw bar (p, 2mm,  90)
  enddef;}
                \fmftop{i1}
                \fmfbottom{s1}
                \fmffixed{(-.1h,.3h)}{v3,v1}
                \fmffixed{(.1h,.3h)}{v3,v2}
                \fmf{tfermion}{i1,v1}
                \fmf{tfermion,tension=.4}{v1,v3,v2}
                \fmf{plain,tension=.4}{v1,v2}
                \fmf{plain}{v3,s1}
                \fmfdot{v1,v2,v3}
             \end{fmfgraph*}
        \end{tabular}
        \end{fmffile}}  \longrightarrow
\parbox{40mm}{\begin{fmffile}{ex4a}
        \begin{tabular}{c}
            \begin{fmfgraph*}(30,40)
       \fmfcmd{
   vardef bar (expr p, len, ang) =
    ((-len/2,0)--(len/2,0))
       rotated (ang + angle
         direction length(p)/2 of p)
       shifted point length(p)/2 of p
   enddef;
   style_def tfermion expr p =
    draw p;
   ccutdraw bar (p, 2mm,  90)
  enddef;}
                \fmftop{i1}
                \fmfbottom{s1}
                \fmf{tfermion}{i1,v1}
                \fmffixed{(0,.3h)}{v2,v1}
                \fmf{tfermion,left=.6,tension=.3}{v1,v2}
                \fmf{tfermion,right=.6,tension=.3}{v1,v2}
                \fmf{plain}{v2,s1}
                \fmfdot{v1,v2}
             \end{fmfgraph*}
        \end{tabular}
        \end{fmffile}}
\fe
in which case we need an extra factor $-\frac{1}{2}$ to ``undo the trace".

\section{The 4-loop correction to the fixed point locus}

In this setion we calculate the coefficient $a_1$ in (\ref{mucorr}). The wave function renormalization that contributes to the second term on the RHS of (\ref{mucorr}) is evaluated by the supergraph

\bea
\parbox{25mm}{\begin{fmffile}{WFR}
        \begin{tabular}{c}
            \begin{fmfgraph*}(45,25)
                \fmfstraight
                \fmfset{arrow_len}{.3cm}\fmfset{arrow_ang}{12}
                \fmfleft{i1}
                \fmfright{o1}
                \fmffixed{(.2h,.4h)}{v1,v2}
                \fmffixed{(-.2h,.4h)}{v4,v3}
                \fmf{fermion}{i1,v1}
                \fmf{fermion,tension=.5}{v4,v1}
                \fmf{fermion}{v4,o1}
                \fmf{fermion,right=.5,tension=0}{v2,v1}
                \fmf{fermion,left=.5,tension=0}{v2,v1}
                \fmf{fermion,right=.5,tension=0}{v2,v3}
                \fmf{fermion,left=.5,tension=0}{v2,v3}
                \fmf{fermion,right=.5,tension=0}{v4,v3}
                \fmf{fermion,left=.5,tension=0}{v4,v3}
            \end{fmfgraph*}
        \end{tabular}
       \end{fmffile}}
\eea
or in component fields, the sum of two graphs
\ie
&W_1 = \parbox{45mm}{\begin{fmffile}{W1}
        \begin{tabular}{c}
            \begin{fmfgraph*}(45,25)
   \fmfcmd{
   vardef bar (expr p, len, ang) =
    ((-len/2,0)--(len/2,0))
       rotated (ang + angle
         direction length(p)/2 of p)
       shifted point length(p)/2 of p
   enddef;
   style_def tfermion expr p =
    draw p;
   ccutdraw bar (p, 2mm,  90)
  enddef;}
                \fmfleft{i1}
                \fmfright{o1}
                \fmffixed{(.2h,.4h)}{v1,v2}
                \fmffixed{(-.2h,.4h)}{v4,v3}
                \fmf{plain}{i1,v1}
                \fmf{plain,tension=.5}{v4,v1}
                \fmf{plain}{v4,o1}
                \fmf{tfermion,right=.5,tension=0}{v2,v1}
                \fmf{tfermion,left=.5,tension=0}{v2,v1}
                \fmf{plain,right=.5,tension=0}{v2,v3}
                \fmf{plain,left=.5,tension=0}{v2,v3}
                \fmf{tfermion,right=.5,tension=0}{v4,v3}
                \fmf{tfermion,left=.5,tension=0}{v4,v3}
            \end{fmfgraph*}
        \end{tabular}
       \end{fmffile}} = {A_{ss}}(2,2){A_{ff}}(2,2)^2{A_{ss}}(2,8-3d) p^{4d-10},
\\ &
W_2 = \parbox{45mm}{\begin{fmffile}{W2}
        \begin{tabular}{c}
            \begin{fmfgraph*}(45,25)
   \fmfcmd{
   vardef bar (expr p, len, ang) =
    ((-len/2,0)--(len/2,0))
       rotated (ang + angle
         direction length(p)/2 of p)
       shifted point length(p)/2 of p
   enddef;
   style_def tfermion expr p =
    draw p;
   ccutdraw bar (p, 2mm,  90)
  enddef;}
                \fmfleft{i1}
                \fmfright{o1}
                \fmffixed{(.2h,.4h)}{v1,v2}
                \fmffixed{(-.2h,.4h)}{v4,v3}
                \fmf{plain}{i1,v1}
                \fmf{tfermion,tension=.5}{v4,v1}
                \fmf{plain}{v4,o1}
                \fmf{tfermion,right=.5,tension=0}{v2,v1}
                \fmf{plain,left=.5,tension=0}{v2,v1}
                \fmf{plain,right=.5,tension=0}{v2,v3}
                \fmf{tfermion,left=.5,tension=0}{v2,v3}
                \fmf{tfermion,right=.5,tension=0}{v4,v3}
                \fmf{plain,left=.5,tension=0}{v4,v3}
            \end{fmfgraph*}
        \end{tabular}
       \end{fmffile}} = -{A_{fs}}(2,2)^3{A_{ff}}(10-3d,2) p^{4d-10}.
\fe
The total contribution is given by
\bea
W_1+8W_2=-\frac{1}{4096\pi^2 (d-3)}p^2.
\eea
Since we have normalized the coefficient of the quadratic term in the moment map to be unity, to obtain the normalized coefficient $a_1$ we need to divide by the two-loop wave function renormalizaiton,
from the diagram
\bea
W_3 = \parbox{45mm}{\begin{fmffile}{W3}
        \begin{tabular}{c}
            \begin{fmfgraph*}(45,25)
   \fmfcmd{
   vardef bar (expr p, len, ang) =
    ((-len/2,0)--(len/2,0))
       rotated (ang + angle
         direction length(p)/2 of p)
       shifted point length(p)/2 of p
   enddef;
   style_def tfermion expr p =
    draw p;
   ccutdraw bar (p, 2mm,  90)
  enddef;}
                \fmfleft{i1}
                \fmfright{o1}
                \fmf{plain,tension=9}{i1,v1}
                \fmf{tfermion}{v1,v2}
                \fmf{tfermion,left=.7}{v1,v2}
                \fmf{plain,right=.7}{v1,v2}
                \fmf{plain,tension=9}{v2,o1}
            \end{fmfgraph*}
        \end{tabular}
       \end{fmffile}} = {A_{ff}}(2,2){A_{ss}}(2-d,2) p^{2d-4}.
\eea
Taking into account the combinatorical factor, the two-loop wave function renormalization is given by
\bea
3W_3 = \frac{1}{32\pi^2(d-3)}p^2.
\eea
Thus we find the result
\bea
a_1=\frac{W_1+8W_2}{3W_3}=-\frac{1}{128}.
\eea

\section{The 2-loop contribution to the two-point function of chiral primaries}

Now, we calculate the coefficients $a_2$ and $a_3$ in the next-to-leading correction to the Zamolodchikov metric. In fact the term proportional to $a_3$ does not appear in the metric on ${\cal M}$, since it does not involve the two-point function of the chiral primaries. Rather, $a_3$ will be a logarithmically divergent coefficient related to the anomalous dimension of the non-primary chiral quartic operators. We include its computation here for completeness. The relevant supergraphs are

\bigskip

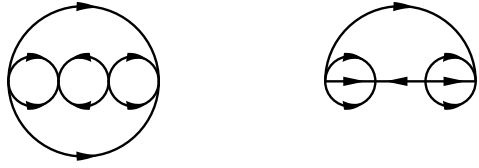
\centerline{\begin{fmffile}{Zam1}
        \begin{tabular}{c}
            \begin{fmfgraph*}(25,25)
                \fmfstraight
                \fmfset{arrow_len}{.3cm}\fmfset{arrow_ang}{12}
                \fmfleft{i1}
                \fmfright{o1}
                \fmf{fermion,right=1}{i1,v1}
                \fmf{fermion,left=1}{i1,v1}
                \fmf{fermion,right=1}{v2,v1}
                \fmf{fermion,left=1}{v2,v1}
                \fmf{fermion,right=1}{v2,o1}
                \fmf{fermion,left=1}{v2,o1}
                \fmf{fermion,right=1}{i1,o1}
                \fmf{fermion,left=1}{i1,o1}
            \end{fmfgraph*}
        \end{tabular}
       \end{fmffile}
~~~~~~~~
\begin{fmffile}{Zam2}
        \begin{tabular}{c}
            \begin{fmfgraph*}(25,25)
                \fmfstraight
                \fmfset{arrow_len}{.3cm}\fmfset{arrow_ang}{12}
                \fmfleft{i1}
                \fmfright{o1}
                \fmf{fermion,right=1}{i1,v1}
                \fmf{fermion}{i1,v1}
                \fmf{fermion,left=1}{i1,v1}
                \fmf{fermion,tension=3}{v2,v1}
                \fmf{fermion,right=1}{v2,o1}
                \fmf{fermion}{v2,o1}
                \fmf{fermion,left=1}{v2,o1}
                \fmf{fermion,left=1}{i1,o1}
            \end{fmfgraph*}
        \end{tabular}
       \end{fmffile}
}

\noindent In above diagrams it is understood that chiral operators of the form ${\rm Tr}\left(\phi^i\phi^j\phi^k\phi^l\right)$ are inserted on the left and right. In component fields, we have two contributions
\ie
Z_1= \parbox{25mm}{\begin{fmffile}{Z1}
        \begin{tabular}{c}
            \begin{fmfgraph*}(25,25)
   \fmfcmd{
   vardef bar (expr p, len, ang) =
    ((-len/2,0)--(len/2,0))
       rotated (ang + angle
         direction length(p)/2 of p)
       shifted point length(p)/2 of p
   enddef;
   style_def tfermion expr p =
    draw p;
   ccutdraw bar (p, 2mm,  90)
  enddef;}
                \fmfleft{i1}
                \fmfright{o1}
                \fmf{plain,right=1}{i1,v1}
                \fmf{plain,left=1}{i1,v1}
                \fmf{tfermion,right=1}{v2,v1}
                \fmf{tfermion,left=1}{v2,v1}
                \fmf{plain,right=1}{v2,o1}
                \fmf{plain,left=1}{v2,o1}
                \fmf{plain,right=1}{i1,o1}
                \fmf{plain,left=1}{i1,o1}
            \end{fmfgraph*}
        \end{tabular}
       \end{fmffile}}  &= {A_{ff}}(2,2){A_{ss}}(2,2)^2 {A_{ss}}(2,10-3d){A_{ss}}(2,12-4d)p^{5d-14}\\
& = \frac{p}{16384\pi^2},
\\
Z_2 = \parbox{25mm}{\begin{fmffile}{Z2}
        \begin{tabular}{c}
            \begin{fmfgraph*}(25,25)
   \fmfcmd{
   vardef bar (expr p, len, ang) =
    ((-len/2,0)--(len/2,0))
       rotated (ang + angle
         direction length(p)/2 of p)
       shifted point length(p)/2 of p
   enddef;
   style_def tfermion expr p =
    draw p;
   ccutdraw bar (p, 2mm,  90)
  enddef;}
                \fmfleft{i1}
                \fmfright{o1}
                \fmffixed{(.3h,0)}{i1,v1}
                \fmffixed{(-.3h,0)}{o1,v2}
                \fmf{plain,right=1}{i1,v1}
                \fmf{plain}{i1,v1}
                \fmf{plain,left=1}{i1,v1}
                \fmf{plain,tension=.5}{v1,v3}
                \fmf{tfermion,tension=1.5}{v3,v4}
                \fmf{tfermion,tension=1.5}{v4,v5}
                \fmf{plain,tension=.5}{v5,v2}
                \fmf{plain,right=1}{v2,o1}
                \fmf{plain}{v2,o1}
                \fmf{plain,left=1}{v2,o1}
                \fmf{plain,left=1}{i1,o1}
            \end{fmfgraph*}
        \end{tabular}
       \end{fmffile}}  &= -{A_{ss}}(2,2)^2 {A_{ss}}(4-d,2)^2 {A_{ss}}(12-4d,2) p^{5d-14} \\
& = -\frac{p}{4096\pi^4 (d-3)}+{\rm finite~terms}.
\fe
Comparing with (\ref{a2}), we need to normalize the above contributions by the tree level two-point function,
\be
\begin{aligned}
Z_3 = \parbox{25mm}{\begin{fmffile}{Z3}
        \begin{tabular}{c}
            \begin{fmfgraph*}(25,25)
                \fmfleft{i1}
                \fmfright{o1}
                \fmf{plain,left=1}{i1,o1}
                \fmf{plain,left=.5}{i1,o1}
                \fmf{plain,right=1}{i1,o1}
                \fmf{plain,right=.5}{i1,o1}
            \end{fmfgraph*}
        \end{tabular}
       \end{fmffile}} & = {A_{ss}}(2,2)^2 {A_{ss}}(4-d,4-d) p^{3d-8} = -\frac{p}{256\pi^2}.
\end{aligned}
\ee
Finally, taking into account combinatorical factors, we obtain the result
\ie
&a_2 = \frac{4Z_1}{Z_3}=-\frac{1}{16},\\
&a_3 = \frac{4Z_2}{Z_3}=\frac{1}{4\pi^2(d-3)}+{\rm finite~terms}.
\fe

\end{document}